\documentclass[sigconf]{acmart} 

\usepackage{diagbox}

\usepackage{mathrsfs}
\usepackage{tikz}
\usepackage{amsmath}
\usepackage{comment}
\usepackage{enumitem}
\usepackage{algorithm}
\usepackage{algpseudocode}
\usepackage{float}
\usepackage{filecontents}
\usepackage{hyperref}
\usepackage{amsmath}
\DeclareMathOperator*{\argmax}{argmax} % thin space, limits underneath in displays
\usepackage{multirow}
\usepackage{url}
\usepackage{framed}
\usepackage{mdframed}
\usepackage{mathtools, nccmath}
\usepackage{balance}
\usepackage{threeparttable}
\usepackage{amsfonts}
\usepackage{mathrsfs}
\usepackage{tikz}
\usepackage{amsmath}
\usepackage{comment}
\usepackage{enumitem}
\usepackage{algorithm}
\usepackage{algpseudocode}
\usepackage{float}
\usepackage{filecontents}
\usepackage{amsmath} 
\usepackage{multirow}
\usepackage{url}
\usepackage{framed}
\usepackage{mdframed}
\mdfsetup{skipabove=3pt,skipbelow=2pt}
 
\usepackage{mathtools, nccmath}
\usepackage[skip=4pt]{caption}
\usepackage{xspace}
\usepackage{booktabs,siunitx,caption}

\newcommand{\sysname}{\emph{Jujutsu}\xspace}

\usepackage{wrapfig}

\usepackage[font=small,labelfont=bf]{caption}

\begin{document}
\raggedbottom
%-------------------------------------------------------------------------------

%don't want date printed
\date{}

% make title bold and 14 pt font (Latex default is non-bold, 16 pt)
%\title{Turning Your Strength against You: Detecting and Mitigating Robust and Universal Adversarial Patch Attacks}

\title{\sysname: A Two-stage Defense against Adversarial Patch Attacks on Deep Neural Networks}

\author{Zitao Chen} 
 
  \affiliation{\institution{University of British Columbia}
      \country{Vancouver, BC, Canada}} 
\email{zitaoc@ece.ubc.ca}

\author{Pritam Dash}  
  \affiliation{\institution{University of British Columbia}
      \country{Vancouver, BC, Canada}}
  \email{pdash@ece.ubc.ca}

\author{Karthik Pattabiraman}
\affiliation{\institution{University of British Columbia}
      \country{Vancouver, BC, Canada}} 
\email{karthikp@ece.ubc.ca}
 
 \copyrightyear{2023}
\acmYear{2023}
\acmConference[Asia CCS'23]{ACM ASIA Conference on Computer and Communications Security}{10--14 July, 2023}{Melbourne, Australia}
\acmBooktitle{The 18th ACM ASIA Conference on Computer and Communications Security (ASIA CCS 2023), 10--14 July, 2023, Melbourne, Australia } 
\acmDOI{xxxxxx}
\acmISBN{xxx}
%-------------------------------------------------------------------------------
\begin{abstract}
%-------------------------------------------------------------------------------

{Adversarial patch attacks} create adversarial examples by injecting arbitrary distortions within a bounded region of the input to fool deep neural networks (DNNs). These attacks are \emph{robust} (i.e., physically-realizable) and \emph{universally} malicious, and hence represent a severe security threat to real-world DNN-based systems.

We propose \sysname, a two-stage technique to detect {and} mitigate robust and universal adversarial patch attacks. 
We first observe that adversarial patches are crafted as localized features that yield large influence on the prediction output, and continue to dominate the prediction on {\em any} input. \sysname leverages this observation for accurate attack detection with low false positives.   
Patch attacks corrupt only a localized region of the input, while the majority of the input remains unperturbed. 
Therefore, \sysname leverages generative adversarial networks (GAN) to perform localized attack recovery by synthesizing the semantic contents of the input that are corrupted by the attacks, and reconstructs a ``clean'' input for correct prediction.

We evaluate \sysname on four diverse datasets spanning 8 different DNN models, and find that it achieves superior performance and significantly outperforms four existing defenses.
{We further evaluate \sysname against physical-world attacks, as well as adaptive attacks. }

\end{abstract}

%%
%% The code below is generated by the tool at http://dl.acm.org/ccs.cfm.
%% Please copy and paste the code instead of the example below.
%%

%%
%% Keywords. The author(s) should pick words that accurately describe
%% the work being presented. Separate the keywords with commas.
\maketitle

\section{Introduction} 

\label{sec:intro}

{
	DNNs are widely used in various application domains, such as autonomous driving~\cite{rao2018deep,bojarski2016end}, facial recognition~\cite{sun2015deepid3,parkhi2015deep} and healthcare~\cite{sahiner2019deep,esteva2019guide}. 
	Unfortunately, DNNs are known to be vulnerable to \emph{adversarial attacks}, which maliciously perturb the inputs  to cause the DNNs to misbehave~\cite{szegedy2013intriguing}.
}
{
	Different variants of adversarial attacks have been proposed in the literature, including \emph{universal} adversarial attacks that cause misclassification on arbitrary inputs~\cite{huang2020universal,li2020advpulse,song2020universal}; and \emph{robust} adversarial attacks that can remain adversarial even when translated to the physical world~\cite{brown2017adversarial,eykholt2018robust,athalye2018synthesizing}.
}

A sub-category of adversarial attacks are {\em adversarial patch attacks} that perform arbitrary changes to the input images within a region of bounded size, in order to cause targeted image misclassification in DNNs~\cite{brown2017adversarial}. These attacks create \emph{robust and universal} adversarial examples - AEs (henceforth referred to as {\em patch attacks}).  
They are an important threat as they entail dire consequences for real-world safety-critical systems such as autonomous vehicles.  
Further, their universal nature drastically lowers the adversary's barrier to launch the attack: an universal adversarial patch can be widely distributed to fool arbitrary DNN systems with little effort.

Patch attacks have been the subject of considerable study, and many techniques have been proposed to detect~\cite{chou2020sentinet,jha2019attribution} and mitigate~\cite{naseer2019local,rao2020adversarial,wu2020defending,hayes2018visible,xiang2020patchguard} them. 
For example, SentiNet~\cite{chou2020sentinet} detects patch attacks based on model interpretability and statistical analysis. LGS~\cite{naseer2019local} mitigates patch attacks by smoothing out the important features in an image based on pre-defined thresholds.
Adversarial training has also been adopted for countering patch attacks~\cite{rao2020adversarial,wu2020defending}.
Unfortunately, these techniques suffer from one or more of the following limitations, 
(1) high false-positives rates (FPR) - unable to correctly distinguish between adversarial and benign image features~\cite{chou2020sentinet,jha2019attribution,naseer2019local,xiang2020patchguard}.
(2) poor detection performance - unable to reliably locate the region of adversarial patch~\cite{chou2020sentinet}.
(3) low mitigation performance (i.e., robust accuracy on adversarial examples) - unable to allow the DNNs to make correct inference on the adversarial examples as many important features are  corrupted~\cite{naseer2019local,rao2020adversarial,wu2020defending,xiang2020patchguard}.

To address the above issues, we propose \sysname\footnote{Jujutsu is a  martial art whose philosophy is to manipulate the opponent's force against him- or herself rather than confronting them with one's own force. Our technique has a similar philosophy, and hence the name. } 
for both detecting {and} mitigating adversarial patch attacks.   
We first outline the challenges in attack detection and attack mitigation, then explain the main ideas to address them.

	\textbf{Attack detection}. 
	The key challenge in accurate attack detection with a low FPR is to identify the unique symptom that characterizes adversarial examples \emph{and} exhibits differences with benign examples. 
	{
		Our solution is based on two insights: (1) the adversarial patch is crafted to constitute localized features in the input, which exerts a large influence on the output in order to manipulate the output; (2) it exhibits the dominant influence on \emph{any} input (input-agnostic).  
		\sysname is built on both insights to expose this behavior of the patch attacks and distinguish AEs from benign examples. 
	}
	
	We leverage the first insight to identify the potential location of the adversarial patch in AEs. 
	Specifically, we propose to locate the adversarial patch region by locating the \emph{salient features} in the saliency map, 
	which are the features that have a large influence on the output (similar to how adversarial patch would behave).   
	However, benign features in  AEs may also have large influence on the output, and hence they may be \emph{conflated} with the adversarial patch (undesirable). 
	To resolve this, we propose a robust method to \emph{pre-process} the saliency map, which can {highlight} the regions associated with the adversarial patch so that \sysname can correctly  identify the region associated with the adversarial patch (instead of benign features). This enables \sysname to reliably locate the adversarial patch from AEs (Section~\ref{sec:detection}). 
	
	We build on the second insight to distinguish the adversarial patch from a benign patch, which is important because the incoming input can be either adversarial or benign, and thus the extracted patch can also be adversarial (i.e., from adversarial examples) or benign (i.e., from benign examples). 
	Specifically, we propose a \emph{guided feature transplantation} method to {strategically} transfer the extracted patch from the original input {to a dedicated region} ({the region where the \emph{least}-salient features reside}) 
	in a new input, and determine whether the patch continues to cause misclassification of the new input.
	If so, it is likely to be an adversarial patch.

	\textbf{Attack mitigation. }
	A natural solution for attack mitigation is to simply mask the entire patch region, and let the DNNs perform inference on the remaining features.
	However, with this approach, the important features in the original images may be corrupted (overridden) by the adversarial patch, and hence it is difficult for the DNNs to make correct predictions on the remaining uncorrupted features (e.g., see Fig.~\ref{fig:mitigation}).
	
	We make the observation that, \emph{patch attacks only perturb a localized region, and hence the majority of image pixels are uncorrupted} (Section~\ref{sec:adv-model}). These pixels can be used to reconstruct the semantic contents in the pixels corrupted by the attacks. Therefore, we use GANs to perform localized attack mitigation, by reconstructing the uncorrupted contents from the corrupted region, which creates the ``clean'' images from AEs for correct prediction.  
	{In addition to improving robust accuracy, our mitigation technique can also be leveraged to further {reduce FPs on benign examples} (Section~\ref{sec:reduce-fp}).}
	
	Finally, different applications may prioritize different defense goals, and hence a configurable defense technique is important. For example, 
	in some systems, the detection performance should be prioritized as an undetected intrusion might cause severe property damage, and hence higher FPR may be acceptable in those settings.  
	For this purpose, we propose a parametric defense strategy that allows for balancing  between detection performance and FPR. 
	We find that the targeted misclassification caused by the adversarial patch often becomes ineffective even \emph{without} completely performing attack recovery on the entire patch, based on which we introduce a parametric attack mitigation strategy (Section~\ref{sec:masking}).

\textbf{Contributions:} The contributions of this work are as follows. 

\begin{itemize}[leftmargin=*]
	\itemsep0em 
	\item A novel patch attack detection method that can reliably locate the regions of adversarial patches in adversarial examples and effectively distinguish between adversarial and benign examples.
	\item A novel attack mitigation technique that leverages the generative power of GANs to allow  the DNNs to make correct predictions on AEs (high robust accuracy), and distinguish false detection on benign examples (low FPRs). It further provides configurability to  balance between  the detection of AEs and FPRs.
	\item A comprehensive evaluation of \sysname on four datasets (ImageNet, ImageNette, CelebA and Place365) spanning eight different DNNs. We find that \sysname achieves superior detection and mitigation performance with low FPRs, and outperforms four existing defenses: LGS~\cite{naseer2019local}, SentiNet~\cite{chou2020sentinet}, adversarial training~\cite{rao2020adversarial,wu2020defending} and PatchGuard~\cite{xiang2020patchguard}.
	\sysname can further defend against both  physical-world attacks and adaptive attacks. 
\end{itemize}

\section{Background}
\label{sec:background}

\subsection{Attack Formulation}
We express a DNN as $F_{\theta} : X \rightarrow Y$, where $X \in \mathbb{R}^n$ and $Y \in \mathbb{R}^m$ denotes the input and output space, and $F$ is parameterized by weights $\theta$ (hereafter omitted for simplicity).  
%Pixel intensity is rescaled from $[0,255]$ to $[0,1]$.
$\bar{y_i}$ is the ground truth label and $\hat{y}=\text{argmax}F_{\theta}(x)$ the prediction label. % with the highest probability.
We call an input $x' \in X$ an adversarial example if 
\begin{equation}
\label{eq:basic-attack}
x' \in X \wedge \argmax{F(x')} = y^{adv} \wedge \argmax{F(x)}=\bar{y}, 
\end{equation}
where $y^{adv}$ is the target class%as we consider targeted misclassification
, $x'$ is the adversarial example generated from the original input $x$.
Patch attack replaces a part of the image with an image patch~\cite{brown2017adversarial}, denoted as $\delta \in \mathbb{R}^n$:
%The generation of $x'$ can be represented as:
\begin{equation}
x' = (1- m) \odot x + m \odot \delta,
\end{equation}

where $m \in \{0,1\}^n$ is a mask used to put the adversarial patch ($\forall m_i \in m, m_i=1$ is where the patch will be placed), $\odot$ is element-wise multiplication, $\delta$ is the adversarial patch.

To make patch $\delta$ be universal (i.e., input-agnostic), the patch is trained over a variety of images. For each input $x \in X$, patch $\delta$ can be applied in any random location $L$.

To make patch $\delta$ robust (i.e., physically realizable), \cite{brown2017adversarial} propose to use a variant of Expectation over Transformation (EOT) framework \cite{athalye2018synthesizing}. 
EOT is used for a distribution of environmental transforms $T$ that transform $x$ to different physical environments (e.g., translation, rotation, lightness changes), under which the adversarial examples aim to remain robust.
Based on the above, the objective function of the patch attack can be formulated as:
\begin{equation}
\label{eq:attack-gen}
\delta = \argmax_{\delta}\mathbb{E}_{x \sim X, t \sim T, l \sim L}[\text{log}\mathrm{Pr}(y=y^{adv}|x') ],
\end{equation} 

where $T$ is a distribution of transformations over the patch, and $L$ is a distribution over locations in the images. 
This 
allows the patch to work regardless of the background.

\subsection{Threat Model}
\label{sec:adv-model}
We assume a white-box attacker with full knowledge of the victim DNN like its structure and parameters. % and training procedure. 
We assume however that the attacker has no  knowledge of the exact inputs to the DNN, but instead has access to a surrogate dataset, which follows the same distribution as the legitimate inputs. 
This is similar to the assumption in other universal attack papers, which have  shown that the knowledge of the input distribution often suffices for the attacker to generate universal adversarial perturbations~\cite{huang2020universal,li2020advpulse,song2020universal}.
{
As in other defense studies~\cite{naseer2019local,xiang2020patchguard,levine2020randomized,mccoyd2020minority}, we consider an adversary who replaces a contiguous region of an image with a { single} adversarial patch - thus, the adversarial patch is localized to a single region of the image  
and the adversary's goal is to {universally} cause targeted misclassification on any input\footnote{Section~\ref{sec:limitation} discusses attack variants (e.g., multi-patch attacks) that are outside our threat model.}.
}
Further, the defender has access to a hold-out set {hidden from the attacker}, which can be created by randomly sampling a series of images from the data distribution.

\section{Methodology}
\label{sec:methodology}
\subsection{Design Overview} 
\label{sec:overview}

\begin{figure}[t]
	\centering
	\includegraphics[ width=3.5in, height=1.4in]{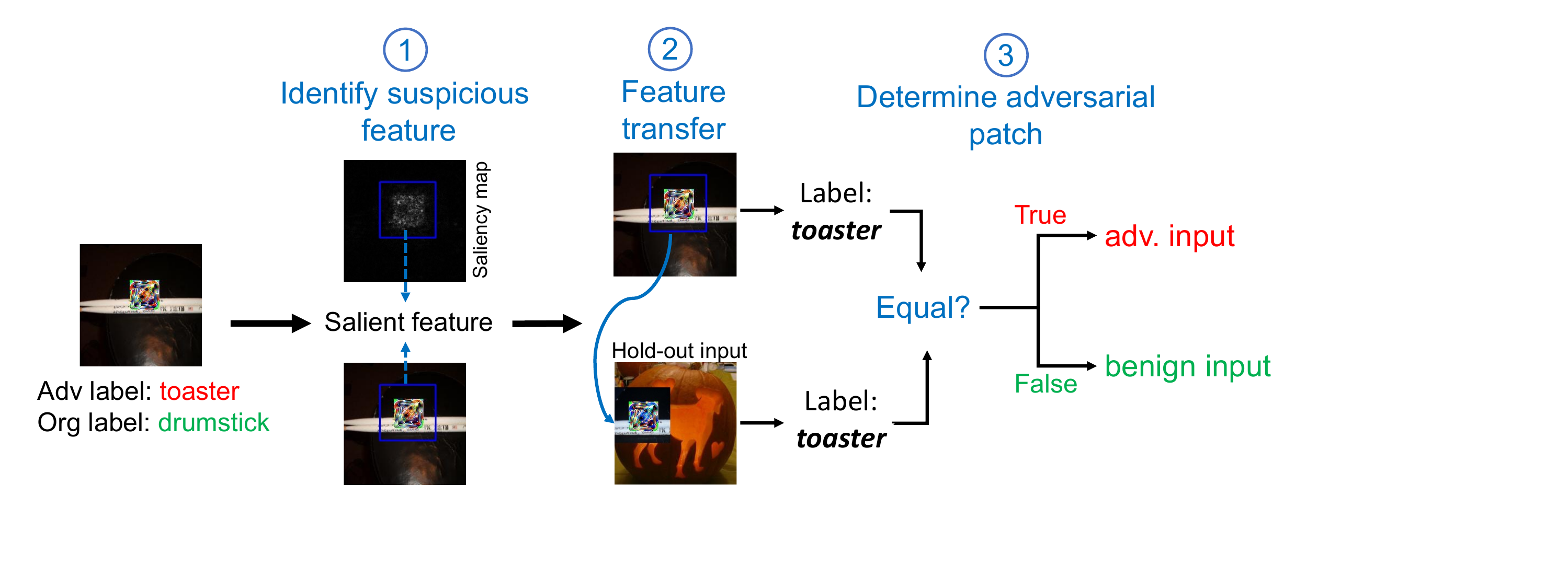}
	\caption{Attack \emph{detection}. Step 1: Identify suspicious features that may contain adversarial patch. Step 2: Transfer the suspicious features to a hold-out input. Step 3: Determine the maliciousness of the suspicious features based on prediction consistency.}
	\label{fig:detection}
	%\vspace{-2mm}
\end{figure}

\begin{figure}[t]
	\centering
	\includegraphics[  height=0.7in]{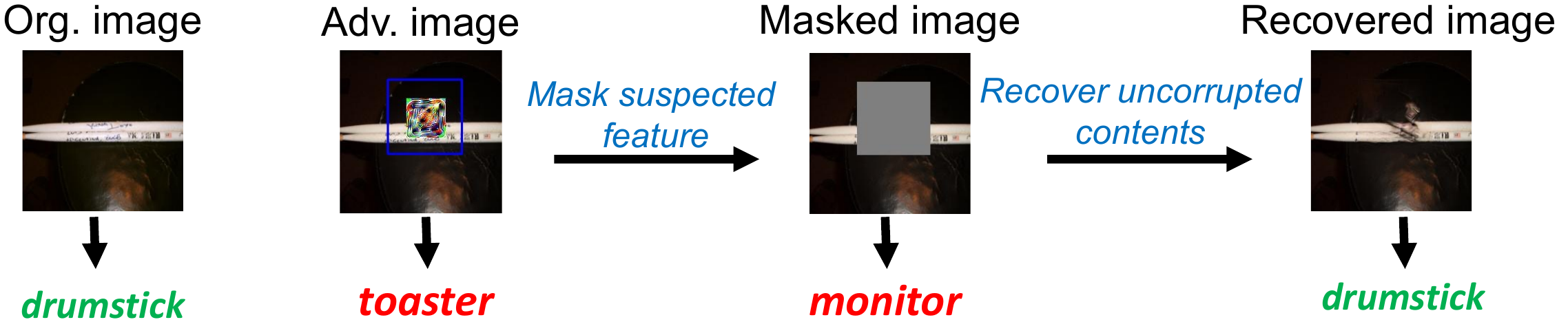}
	\caption{Attack \emph{mitigation}: (randomly) mask the suspicious features (blue box), and use GAN to recover the uncorrupted contents in the mask. }
	\label{fig:mitigation}
	\vspace{-4mm}
\end{figure}

Fig.~\ref{fig:detection} illustrates attack detection and Fig.~\ref{fig:mitigation} attack mitigation.

\textbf{Detecting the adversarial patch.} 
We first identify \emph{suspicious features} that potentially contain the adversarial patch. 
We observe that the universal and localized nature of adversarial patch induces the perturbations to have a {\em disproportionately large} influence on the output in order to dominate the prediction on {\em any} input, which can be exposed by investigating the \emph{salient features} from the saliency map (Step 1 in Fig.~\ref{fig:detection}). 
These salient features are considered suspicious as they have a large influence on the output, similar to the adversarial patch's behavior. 
However, salient features may also point to the natural features in the images and hence the adversarial patch region may be \emph{undetected} (see the left of Fig.~\ref{fig:saliency_comparison}). 
To avoid this, we \emph{pre-process} the saliency map to highlight the regions that are associated with the adversarial patch, hence we can reliably locate the adversarial patch region (see the right of Fig.~\ref{fig:saliency_comparison}).

On the other hand, since the input can either be adversarial or benign, the suspicious features can also be adversarial (from adversarial example) or benign (from benign example). 
Our idea to distinguish them is based on the observation that the adversarial patch, when transplanted to other images,  \emph{will continue} to trigger misclassification, which is \emph{different} from how benign examples would behave\footnote{{There are two potential scenarios that would lead to false positive on benign samples, and they are discussed in Section~\ref{sec:feature-trans} and Section~\ref{sec:reduce-fp}, respectively.}}.
Therefore, Step 2 extracts the suspicious features from the original input, and transplant them \emph{to the least-salient feature region} ({the region where the \emph{least}-salient features reside}) of hold-out input. %  (we'll also show later randomly transplanting the features is undesirable). 
Step 3 compares the prediction on the original input and the hold-out input implanted with suspicious features. If both predictions lead to the \emph{same} prediction label, the suspicious features are marked as adversarial.

\textbf{Mitigating the adversarial patch.} 
The goal of mitigation is to remove the attacks' effects, and allow the DNN to predict the correct label from the adversarial examples.  
A straightforward solution is to mask out the suspicious features so that the adversarial patch will not contribute to the final prediction. 
Unfortunately, masking alone does not work in many situations. For instance, in Fig.~\ref{fig:mitigation},  masking the suspected feature mitigates the adversarial attack, as the DNN no longer predicts the adversarial example as a ``toaster'' (the target label determined by the attacker), thus defying the attack. 
However, the DNN  predicts the image with the mask as a  ``monitor'', which is clearly not the correct label for the image. 
This shows that merely masking the suspected feature is not sufficient, as it also  removes the semantic contents in the image, and hence the DNN is \emph{unable} to predict the correct label from the masked images. 

{
Our goal is to remove the effects of the attacks while preserving the semantic contents. 
We observe that the adversarial patch is confined within a \emph{localized} region
,  and the majority of the pixels are uncorrupted, which can be used as a rich context to reconstruct the semantic contents corrupted by the attacks. % for attack mitigation. 
}
Specifically, we use generative adversarial networks (GAN)~\cite{yu2019free,liu2018image,zheng2019pluralistic} to reconstruct the semantic contents from the pixels that are masked, resulting in a ``clean'' image that is free from corruptions for the DNN to make correct prediction. As shown in Fig.~\ref{fig:mitigation}, after recovering the contents from the masked regions, the DNN is able to correctly predict the adversarial image as a ``drumstick''. 
We use the prediction label of the recovered image as the final output.

{
Section~\ref{sec:reduce-fp} discusses how our mitigation technique can also be leveraged to  {reduce false detection on benign examples}..
}

\subsection{Detecting the Adversarial Patch}
\label{sec:detection}
\subsubsection{Robust Suspicious Feature Detection.}
\label{sec:feature-det}
We first compute a saliency map that models the contributions of different pixels on the final decision.
One common approach is to compute the gradients of the output with respect to the input pixels. 
Mathematically, the saliency map $M_j(x)$ can be expressed as: $M_j(x) = \partial F(x)_j / \partial x$, where $j$ indicates the class label. 
$M_j(x)$ represents how much difference a tiny change in $x$ would contribute to the output $F(x)_j$. Thus $M_j(x)$ can highlight the key regions in predicting $F(x)_j$. 

We use SmoothGrad~\cite{smilkov2017smoothgrad}, which can visually sharpen the gradient-based saliency map and smooth out the noisy gradients (that arise due to the local variations in partial derivatives~\cite{smilkov2017smoothgrad}). Other methods such as Grad-cam~\cite{selvaraju2017grad}, Integrated Gradient~\cite{sundararajan2017axiomatic} may also be used.
Given the noisy (fluctuating) gradients, SmoothGrad computes a local average of the gradient values, by taking random examples in the neighborhood of an input $x$, and averaging the resulting saliency maps. This operation can be expressed mathematically as:

\begin{equation}
\label{eq:smoothgrad}
\hat{M}_j(x) = \frac{1}{n} \sum_{1}^{n} M_j(x + \mathcal{N}(0, \sigma^2)),
\end{equation}

where $n$ is the number of examples, and $\mathcal{N}(0, \sigma^2)$ represents the Gaussian noise with standard deviation $\sigma$.

\begin{figure}[t]
\centering
  \includegraphics[ height=0.7in]{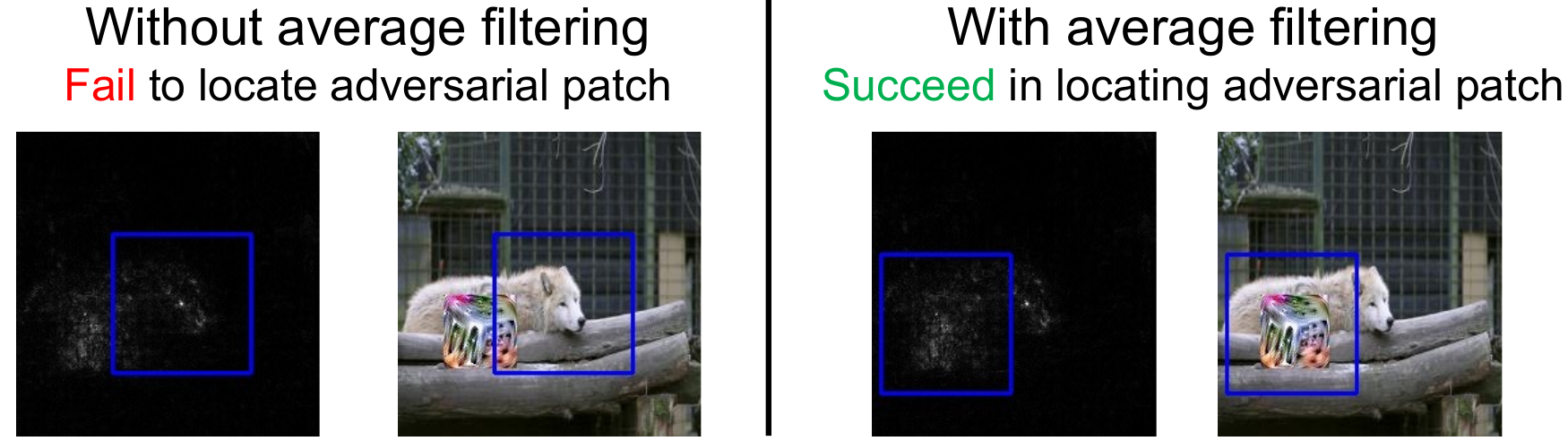}
\caption{Extracting suspicious features from saliency map with and without average filtering. The proposed use of average filtering is able to locate the adversarial patch correctly. }
\label{fig:saliency_comparison}
\vspace{-4mm}
\end{figure}

To extract the suspicious features from the saliency map, we first choose the point that has the maximum value in the saliency map, and draw a detection box around it. 
However, this approach is susceptible to noise and a single large-value pixel outside the adversarial patch could result in a mis-identification. Thus, the detection box would fail to locate the adversarial patch.
{
The reason is that there are many benign features that are uncorrupted in the adversarial examples, which may also have large influence on the outputs, and might be conflated with those of the adversarial patch. 
In the left-hand side of Fig.~\ref{fig:saliency_comparison}, the region associated with the benign feature is identified as the salient feature.
}

Therefore, we perform an \emph{average filtering}~\cite{avg-filtering} to pre-process the saliency map in order to highlight the regions of the adversarial patch, and downplay those of the benign feature.  
It takes the average of all the pixels under the kernel area (in the saliency map) and uses the average value to replace the central element. 
Our  intuition is that the region of adversarial patch has a higher density than that of the benign feature (because it needs to have a disproportionately large influence on the output to dominate the prediction on any input), and hence by performing average filtering, the adversarial patch will remain salient while the benign feature will become less salient, thereby allowing us to accurately locate the adversarial patch.
A visual comparison of the two approaches in identifying the suspicious features is shown in Fig.~\ref{fig:saliency_comparison}.
{Our ablation study (Section~\ref{sec:ablation-study}) also validates that the proposed pre-processing method enables \sysname to detect much more AEs than without it.}

\subsubsection{Guided Feature Transplantation.}
\label{sec:feature-trans}
\begin{figure}[t]
\centering
  \includegraphics[ height=1.2in, width=3.3in]{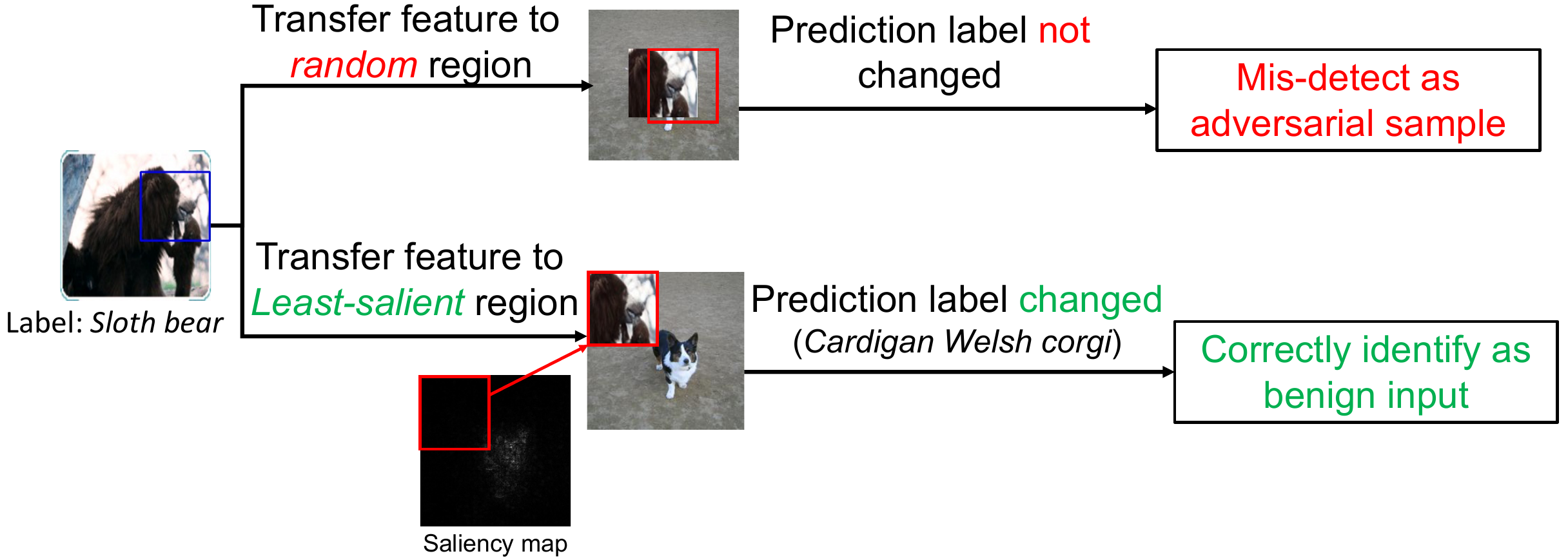}
\caption{Different strategies to transfer features. 
In the top, features are transplanted to a \emph{random} location of the hold-out input which leads to a FP, while those in the bottom are to the \emph{least-salient} region of the hold-out input (our approach), thus avoiding a FP.}
\label{fig:feature-trans} 
\vspace{-2mm}
\end{figure}

{As mentioned, identifying the suspicious features by itself is not enough to determine whether the features are coming from adversarial or benign examples.}
Hence, we transfer the suspicious features from the original input to the hold-out input in order to determine whether they are truly malicious.
One way is to \emph{randomly} transplant the suspicious features to the new hold-out input and compare the prediction. 
However, this may occlude the foreground object in the hold-out input.
Should this happen, the prediction labels on the original and hold-out inputs may become the same, leading to a mis-detection of benign input, i.e., False Positive (FP).
Fig.~\ref{fig:feature-trans} shows an example where randomly transplanting the benign features to a  hold-out input leads to the same prediction label `Sloth bear'', thus resulting in an %mis-detection of the benign image.
FP. 

To avoid FP, we propose a guided feature transplantation method to transplant the suspicious features to the \emph{least-salient} regions of the hold-out input, in order to  minimize the chances that the suspicious features override the hold-out input's foreground object. 
The least-salient regions are those regions that have \emph{low} influence on the output according to the saliency map. 
Only those suspicious features containing the adversarial patch at the least-salient regions will also lead to the same prediction label (due to the patch's universal nature).  
Fig.~\ref{fig:feature-trans} shows how this method works and
{
our ablation study (Section~\ref{sec:ablation-study}) shows that it is able to yield much lower FPRs (compared with random transplantation). 
}

\subsubsection{Prediction Comparison for Attack Detection.}
\label{sec:pred-comp}
The final step is to compare the prediction labels on the original and hold-out images implanted with suspicious features. 
The original image is deemed to be  adversarial if and only if both images yield the same prediction label. This is because only the suspicious features that contain the adversarial patch will cause (the same) misclassification on the hold-out input. 
We are also able to identify suspicious features that come from the benign features, by checking whether the prediction labels on the original and hold-out images implanted with suspicious features are \emph{different}.
We consider an image to be benign if the prediction labels on the original and hold-out images are different.

\subsection{Mitigating the Adversarial Patch}
\label{sec:mitigation}

\subsubsection{GAN-based Localized Attack Mitigation}
\label{sec:inpaint} 
A natural solution to mitigate the attacks is to mask the adversarial patch, and let the DNNs perform inference on the remaining features.
However, some of the important features in the original images might have been corrupted (overridden) by the adversarial patch, and hence performing masking alone will result in the loss of semantic contents that are crucial for the DNNs to classify the images correctly.

On the other hand, we also note that patch attacks only perturb a small localized region and a large portion of image pixels are intact, which can serve as the rich context to synthesize the contents that are corrupted (replaced) by the attacks.
Based on this observation, we propose to use generative adversarial networks (GAN)~\cite{yu2019free,liu2018image,zheng2019pluralistic,pathak2016context} to perform localized attack mitigation by reconstructing the contents replaced by the adversarial patch, and to increase the probability that the DNN predicts the correct label. 
While there are many GAN techniques proposed in the literature. We use 
PICnet~\cite{zheng2019pluralistic}, a recent technique that can generate multiple and diverse plausible contents from the masked regions. 
Although we choose PICnet in this work, other techniques~\cite{yu2019free,liu2018image,pathak2016context} may also be considered for the same. 

Formally, let $x$ be the original image, $x_{m}$ the image with a region of pixels being masked, and $x_{c}$ the original pixels that are masked. 
PICnet synthesizes diverse contents from the mask by sampling a conditional distribution $p(x_c | x_m)$. 
In the training phase, PICnet uses a \emph{reconstructive} pipeline, in which the missing regions $x_c$ are encoded into the latent space representation in a continuous distribution that can be exampled to rebuild the diverse and plausible $x_c$.
The reconstructive pipeline leverages $x_c$ and $x_m$ to reconstruct $x$ in a supervised manner ($x_c$ is the ground truth). 
In the testing phase, PICnet uses a \emph{generative} pipeline to infer the conditional distribution of $p(x_c | x_m)$, which is exampled to generate $x_c$. 
The parameters in the reconstructive pipeline are shared with the generative pipeline so that it can  reconstruct $x$ from $x_m$ during testing. 
The resulting recovered images are meant to be free from adversarial perturbations, and thus we use the labels on the recovered images as the final output for mitigation.

\subsubsection{Parametric Attack Mitigation}
\label{sec:masking} 
We now introduce our parametric mitigation strategy that allows balancing between the detection performance and FPRs.
The motivation is that it is often \emph{unnecessary} to mask all the pixels in order to make the target misclassification ineffective, e.g., we find masking 75\% of the suspected features is able to change the targeted miclassification in over 99.9\% of the adversarial examples.  

By partially masking suspected features, \sysname allows the defender to reduce the FPR - we explain the reason below.
We can determine a mis-detection on the benign input (i.e., FP) if the predictions on both the original and recovered images result in the same label (to be discussed in Section~\ref{sec:reduce-fp}).  
The fewer pixels that are masked, the better is the quality of the resulting recovered image, because more semantic information is preserved in the image. 

\begin{figure}[t]
\centering
  \includegraphics[ height=1.3in ]{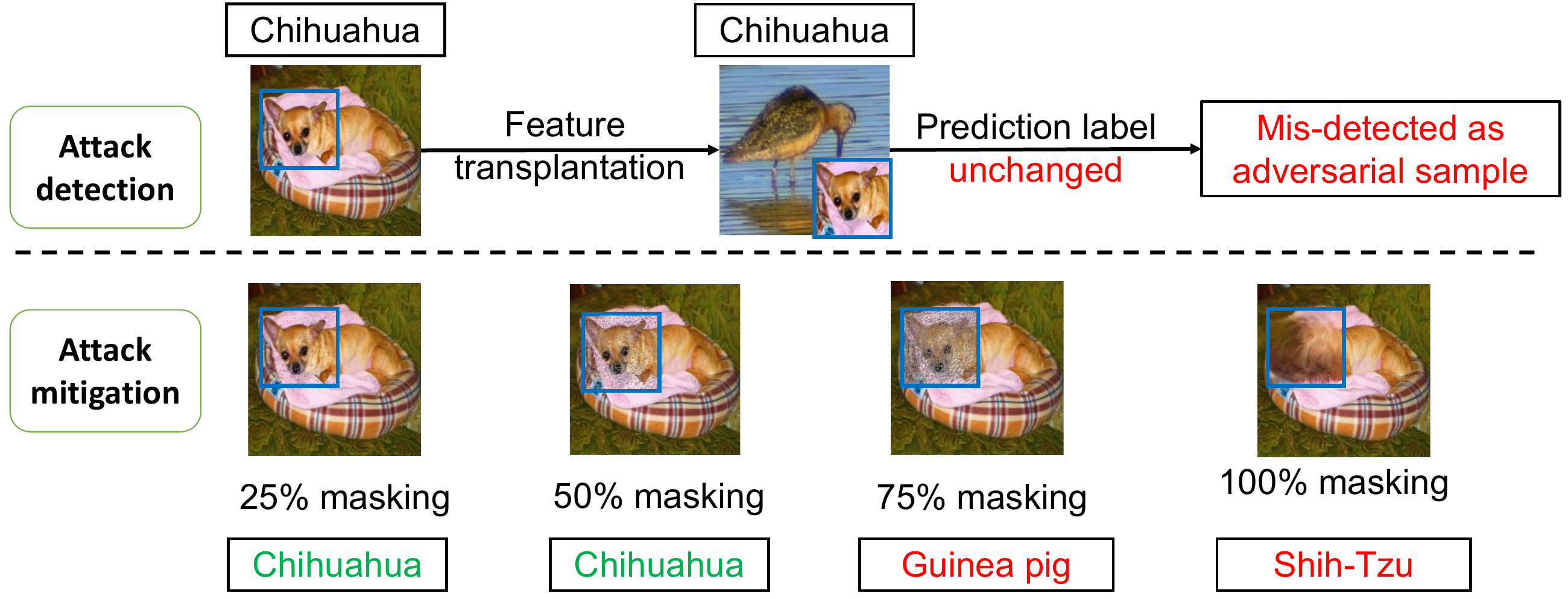}
\caption{Upper row: An illustration of a false detection on benign sample. Lower row: How \sysname may reduce FP in its attack mitigation phase (under different masking percentages).}
\label{fig:second-fp}
\vspace{-4mm}
\end{figure}

Fig.~\ref{fig:second-fp} shows an example of the recovered images under different masking percentages. 
The original Chihuahua image is mis-detected as adversarial, which can be eliminated if both  the original and recovered image have the same label. 
In this example, if 25\% or 50\% of the pixels are masked, \sysname is able to rectify the mis-detection. 
However, if 75\% or 100\% of the suspicious features are masked, the DNN is unable to generate the correct prediction on the recovered image, thus resulting in an FP.
This explains why masking the entire set of suspicious features could be undesirable.

\subsubsection{Reducing FPRs}
\label{sec:reduce-fp}
{Section~\ref{sec:feature-trans} explains the first scenario where FPs might occur. We now explain the second scenario where FPs may still arise, and how \sysname can prevent them.

We use Fig.~\ref{fig:second-fp} to illustrate. 
In the attack detection phase, \sysname transplants the Chihuahua object (as suspicious features) to the hold-out input. 
The resulting hold-out input, originally containing a single Dowitcher, is now classified as a Chihuahua by the DNN, which is the same as the original input. This thus results in a FP.
}

To reduce the above FP, we propose to signal a FP when the prediction label on the original input (that \sysname originally detected as adversarial) and the recovered input are \emph{identical}, during the mitigation phase. 
The intuition is that a benign input does not contain an adversarial patch, and hence predictions based  on the original and the recovered images should both result in the same prediction label. 

The above process to reduce the FPR might also flag some adversarial examples as  benign ones, thereby resulting in missed detection.
For example, if the masking percentage is low, the adversarial patch will continue to cause misclassification on the recovered images, based on which \sysname would incorrectly flag the adversarial patch as a benign image patch.
We study how different masking percentages would affect the detection performance  (Table~\ref{tab:mitigation}).

\subsubsection{Algorithm.} Algorithm~\ref{alg:overall} shows the overall algorithm. 
The inputs are the images to be classified and parameters for \sysname.
%The input is the set of features to be classified.   
For each $x_i$, the output includes the prediction label $y_{x_i}$ and a flag $isAdv_{x_i}$ on whether $x_i$ is adversarial. 
Lines 4-8 extract the salient features from $x_i$. 
Lines 10-15 identify the least-salient regions in the hold-out input $x^*$, which will be replaced by the salient features from $x_i$.
Lines 16-25 perform feature transfer and compare the prediction labels on the original and implanted images. %implanted with salient features.
Lines 29-39 perform attack mitigation by accepting the label from the recovered image, 
and checking for mis-detection (thereby reducing FPs).

\begin{algorithm}[t] 
%\normalsize
%\footnotesize
\scriptsize

\begin{flushleft}
\hspace*{\algorithmicindent} \textbf{Input:} $X_{test}$: Test images; $X_{hold}$: Hold-out images; $F$: DNN model; \\ 
\hspace*{1.0cm} $l$: Length of detection box; $p$: Percentage of pixels to mask \\
\hspace*{\algorithmicindent} \textbf{Output:} 
	$Y_{test}$: Prediction on $X_{test}$; $isAdv_{X_{test}}$: whether $X_{test}$ is adversarial
\end{flushleft} 
  \begin{algorithmic}[1]
    \Function{Detection}{$X_{test}, X_{hold}, F, l$}
    \For{\textbf{each}  $(x_i, y_{x_i}, isAdv_{x_i}) \in (X_{test}, Y_{test}, isAdv_{X_{test}})$ }
    	\State $y_j = \text{argmax}F(x_i)$

    	\State \textbf{// Extract the salient features $B_{x_i}$ from $x_i$}
    	\State $M_j(x_i) = \text{SmoothGrad}(x_i, y_j)$ // Saliency map for $(x_i, y_j)$
    	\State $M_j(x_i) = \text{AverageFilter}(M_j(x_i))$ // Average filtering over saliency map
    	\State $(x_{max}, y_{max}) = \text{MaxLoc}(M_j(x_i))$ // point with maximal value 
   		\State Draw a box $B_{x_i}$ around $(x_{max}, y_{max})$ with length $l$ // suspicious features

   		\State \textbf{// Identify the least-salient features $B_{x^*}$ from $x^*$}
    	\State Randomly select $x^* \in X_{hold}$
    	\State $y^*_k = \text{argmax}F(x^*)$
    	\State $M_k(x^*) = \text{SmoothGrad}(x^*, y^*_k)$ // Saliency map for $(x^*, y^*_k)$
    	\State $M_k(x^*) = \text{AverageFilter}(M_k(x^*))$ // Average filtering over saliency map
    	\State $(x^*_{min}, y^*_{min}) = \text{MinLoc}(M_j(x_i))$ // point with minimal value 
    	\State Draw a box $B_{x^*}$ around $(x^*_{min}, y^*_{min})$ with length $l$ 

    	\State \textbf{// Feature transfer and prediction comparison}
    	\State $x^{**} = x^*.\text{replace}(B_{x^*}, B_{x_i})$ 
    	\State $y^{**}_k = \text{argmax}F(x^{**})$
    	\If{ $y^{**}_k == y_j $ }  
    		\State $y_{x_i}, isAdv_{x_i}$ = \textbf{MITIGATION}($x_i, B_{x_i}, F, p$) // \textbf{$x_i$ is adversarial} 

    	\Else 
    		\State $y_{x_i} = y_j$ // $y_j$ is the prediction label from line 3
    		\State $isAdv_{x_i} = \text{False} $   // \textbf{$x_i$ is benign} 
    	\EndIf
    \EndFor

   	\State \textbf{return} $Y_{test}, isAdv_{X_{test}}$
    \EndFunction
    \\

    \Function{Mitigation}{$x, B_{x}, F, p$}
    	\State $y_{org} = \text{argmax}F(x)$
    	\State $x_{mask}$ = Randomly mask $p$\% of pixels within $B_{x}$ in $x$ 
    	\State $x_{recovered}$ = PICNet($x_{mask}$) // GAN-based recovery
    	\State $y_{new} = \text{argmax}F(x_{recovered})$
    	\If{$y_{org} != y_{new}$}
    		\State \textbf{return} $y_{new}, $ True // \textbf{Attack mitigation}
    	\Else
    		\State \textbf{return} $y_{org}$, False // \textbf{Reduce false positive}
    	\EndIf 
    \EndFunction

  \end{algorithmic}
    \caption{Detect and mitigate patch attacks}
    \label{alg:overall}
    
\end{algorithm}
\vspace{-2mm}

\section{Evaluation}
\label{sec:evaluation}
We first describe the experimental setup of \sysname, and then answer the following research questions (RQs) in subsequent sections.
 
\noindent
\textbf{RQ1:} What's the detection performance of \sysname? 

\noindent
\textbf{RQ2:} What's the mitigation performance of \sysname? 

\noindent
\textbf{RQ3:} How does \sysname compare with existing techniques? 

\noindent
\textbf{RQ4:} Can \sysname defend against physical-world attacks? 

\noindent
\textbf{RQ5:} Can \sysname defend against attacks targeting different classes? 

\noindent
\textbf{RQ6:} Is \sysname able to thwart the adaptive attackers?

\subsection{Experimental Setup}
\label{sec:setup}

\subsubsection{Datasets and Architectures} 
{We evaluate \sysname on ImageNet~\cite{deng2009imagenet}, ImageNette~\cite{imagenette}, CelebA~\cite{liu2015faceattributes} and Place365~\cite{zhou2017places}.
ImageNet is a 1000-class dataset and we use a ResNet-50. 
ImageNette is a 10-class subset of ImageNet and we use a ResNet-18. 
CelebA is a facial dataset with diverse faces. We created a 307-classes subset from the original set and train a ResNet-18 model following \cite{train-celeb} to perform identity classification.
Place365 is a 365-class dataset containing common natural sceneries (e.g., patio, restaurant) and we use a ResNet-50. 
{We evaluate \sysname on 6 more DNN models in Section~\ref{sec:more-dnn}.}

\subsubsection{Attack Setup}
\label{sec:attack-setup}
{ 
The attacker's goal is to synthesize adversarial patches that achieve { high attack success rate}. 
As in prior work~\cite{naseer2019local,xiang2020patchguard,levine2020randomized,mccoyd2020minority}, we consider a square digital patch (we use a circle patch in evaluating physical world attack in Section~\ref{sec:physical-attack}), and discussion on other patch shapes is in Section~\ref{sec:limitation}.
For each dataset, we generate patches of different sizes, occupying 5\%, 6\% and 7\% of the pixels. {The patch is overlaied to a random position in the image.}
We use $x\%$ patch to refer to a patch that occupies $x\%$ of the pixels of the image. 
We do not consider patches of smaller size because we find that they are unable to universally cause misclassification, e.g., use of a 4\% patch on CelebA degraded the attack success rate by more than 45\%. 
{We consider 7\% as the largest patch size because it is already able to achieve very high attack success rate (average over 99\%), and we evaluate \sysname on larger patch sizes (8\%$\sim$10\%) in Section~\ref{sec:larger-patch-eval} for completeness.}
For each patch, we train it for 30 epochs on a training set with 2000 images. We evaluate the attack success rate on a separate test set, and choose the one with the highest success rate. 
For the attack evaluation on ImageNette and CelebA, we use the entire test set in each dataset; for ImageNet and Place365, we use 10000 images from the validation set for each.   
Examples of adversarial examples  can be found in Fig.~\ref{fig:sample-img} (Appendix). 
}

\subsubsection{Defense Setup}
\label{sec:technique-setup}

There are three parameters in our defense setup: (1) kernel size for pre-processing (average filtering) the saliency map; (2) size of the detection bounding box; and (3) number of hold-out images used for feature transplantation. 

We vary \emph{each} parameter under different values and empirically select the one that strikes a good balance between detection performance and FPRs (e.g., a larger detection bounding box may achieve higher detection performance but with higher FPRs) - a detailed evaluation for each parameter under different values is in Appendix~\ref{sec:select-parameters}, based on which we choose a kernel size of 51, a bounding box occupying $\sim$20\% of the image pixels and 2 random hold-out images (out of 1000) for feature transplantation (on all dataset)\footnote{Our code is publicly available at \url{https://github.com/DependableSystemsLab/Jujutsu}.}.

\begin{table}[t]
\caption{Detection performance in terms of detection success recall on adversarial examples (AEs) and detection FPR.}
\label{tab:detection}
\centering 
%\normalsize
\footnotesize
%\scriptsize
\begin{threeparttable}

\renewcommand{\arraystretch}{1}

\begin{tabular}{|l|c|cccc|}
\hline
\multirow{2}{3em}{\bf Dataset} & \multirow{2}{2em}{\bf Patch Size } &    \multirow{2}{3em}{\bf Clean Acy. } &  \multirow{2}{6em}{\bf Attack Success Rate } &  \multirow{2}{7em}{\bf Detection Success Recall} &  \multirow{2}{4em}{\bf Detection FPR\tnote{1} }    \\
                & & & & & 
\\                
\hline
\hline
\multirow{3}{4em}{ImageNet}     & 7\% & \multirow{3}{*}{74.17\%}  & 99.81\% & 99.74\% & 4.24\%  \\
                                & 6\% &                           & 98.34\% & 97.70\% & 4.01\%     \\
                                & 5\% &                           & 94.06\% & 93.37\% & 3.81\%         \\
\hline
\multirow{3}{4em}{ImageNette}   & 7\% & \multirow{3}{*}{98.32\%}  &100.00\% & 100.00\% & 8.11\%  \\
                                & 6\% &                           &99.00\%  & 99.94\% & 9.18\% \\
                                & 5\% &                           &98.36\%  & 99.88\% & 8.98\% \\
                      
\hline
\multirow{3}{4em}{CelebA}      & 7\% & \multirow{3}{*}{83.13\%}  &99.40\% & 99.10\% & 0.00\%  \\
                                & 6\% &                           &96.62\% & 95.16\% & 0.00\% \\
                                & 5\% &                           &83.22\% & 69.09\% & 0.00\% \\
\hline
\multirow{3}{4em}{Place365}       & 7\% & \multirow{3}{*}{54.45\%}  &96.77\% & 98.93\% & 0.57\%  \\
                                & 6\% &                           &96.90\% & 99.29\% & 0.46\% \\
                                & 5\% &                           &97.11\% & 98.97\% & 0.53\% \\   
\hline
\multirow{1}{4em}{\textbf{Average}} & N/A & N/A & 96.63\% & 95.93\% & 3.33\% \\
\hline

\end{tabular}

\begin{tablenotes}
        \footnotesize 
        \item[1] The FPR can be further \emph{reduced} as explained in Section~\ref{sec:reduce-fp} - see Table~\ref{tab:mitigation}.
\end{tablenotes}

\end{threeparttable}
\vspace{-4mm}
\end{table} 
\subsection{RQ1 - Detecting Adversarial Patch Attacks}
\label{sec:detection-eval} 
\textbf{Metrics.} We use \emph{detection success recall} to denote the fraction of adversarial examples detected by \sysname, and \emph{detection FPR} for the fraction of false positives on benign inputs.
Benign inputs are the same as the adversarial inputs, except that they do not have the adversarial patch. 
We consider an image as adversarial if and only if the predicted label for it is identical to that of {\em both the hold-out images} implanted with the suspicious features. %Otherwise, we consider it to be benign. 

Table~\ref{tab:detection} shows \sysname's detection performance on all 4 datasets.

{
\emph{Detection success recall.} \sysname is able to consistently detect adversarial examples, with a detection success recall rate of over 93\% across patch sizes (in most cases).  
The detection success recall increases with the size of the patch as a larger patch has a higher attack success rate. 
On average, \sysname can detect around 96\% of the adversarial examples on all the datasets.
}

{
\emph{Detection FPR.} \sysname yields an average FPR of 3.3\% on the 4 datasets.
We find that the FPRs on the two object-recognition datasets (ImageNet and ImageNette) are higher than that on the facial and scenery datasets. This is because the salient features in object-recognition datasets might contain the entire object (e.g., a small bird), which can cause the model to continue to assign the same label to the transplanted image. 
However, for the facial and scenery datasets, the salient features only contain a fraction of the image pixels (e.g., a partial face), and are hence  unlikely to result in the same label on the transplanted image. 
}

{
Despite the higher FPRs on ImageNet and ImageNette, \sysname's mitigation mechanism is able to further reduce FPRs as explained in  Section~\ref{sec:reduce-fp}. We evaluate the FPR reduction in the next section. 
}

\subsection{RQ2 - Mitigating Adversarial Patch Attacks}
\label{sec:mitigation-eval}

\begin{table*}[t]
\caption{Mitigation performance for: 1) GAN-based recovery and 2) masking-alone recovery.
Better results are marked in \textbf{bold}.}
\label{tab:mitigation}
\centering 
%\normalsize
\footnotesize
\begin{tabular}{|c|l |cccc | cccc| cccc | cccc|}
\hline
\multirow{3}{6.5em}{\textbf{Metric} (\%)} & \multirow{3}{4em}{\textbf{Approach}} &  \multicolumn{4}{|c|}{\textbf{ImageNet}} & \multicolumn{4}{|c|}{\textbf{ImageNette}} & \multicolumn{4}{|c|}{\textbf{CelebA}} & \multicolumn{4}{|c|}{\textbf{Place365}}  \\
                          \cline{3-18}      & &                                  \multicolumn{4}{|c|}{\textbf{Masking percentage}} & \multicolumn{4}{|c|}{\textbf{Masking percentage}} & \multicolumn{4}{|c|}{\textbf{Masking percentage}} & \multicolumn{4}{|c|}{\textbf{Masking percentage}}  \\

  & & 25\% & 50\% & 75\% & 100\% & 25\% & 50\% & 75\% & 100\%  & 25\% & 50\% & 75\% & 100\% & 25\% & 50\% & 75\% & 100\%  \\
\hline

\multirow{2}{6.5em}{\textbf{Robust Accuracy}} & GAN-based  &  41.98 & \textbf{69.70} & \textbf{73.52} & \textbf{77.47} & 21.72 & 85.17 & \textbf{94.73} & \textbf{95.51} & 34.00 & \textbf{51.40} & \textbf{51.81} & \textbf{64.56} & 16.58 &74.03 & \textbf{75.04} & \textbf{81.39} \\
                                            & Masking-alone     &  \textbf{58.15} & 69.32 & 70.48 & 75.24 & \textbf{34.60} & \textbf{90.98} & 94.53 & 94.19 & \textbf{41.91} & 46.00 & 45.27 & 48.12 & \textbf{63.39} & \textbf{74.47} & 74.53 & 75.12 \\

\hline
\hline
\multirow{2}{6.5em}{\textbf{Mitigation FPR}} & GAN-based  &  \textbf{0.34} & \textbf{0.95} & \textbf{1.67} & \textbf{1.74} & \textbf{0.14} & \textbf{0.61} & \textbf{1.06} & \textbf{0.91} & \textbf{0.00} & \textbf{0.00} & \textbf{0.00} & \textbf{0.00} & \textbf{0.00} & \textbf{0.10} & \textbf{0.14} & \textbf{0.20} \\
                                           & Masking-alone     & 1.03 & 1.69 & 2.00 & 1.85 & 0.76 & 1.26 & 1.55 & 1.74 & \textbf{0.00} & \textbf{0.00} & \textbf{0.00} & \textbf{0.00} & 0.08 & 0.17 & 0.22 & 0.25 \\
\hline
\hline
\multirow{2}{6.5em}{\textbf{Mitigation Success Recall}} & GAN-based  &  53.71 & 93.61 & 96.90 & 96.89 & 21.78 & 88.17 &99.38 & 99.70 & 57.60 & 87.33 & 87.75 & \textbf{87.79} & 18.67 & 98.02 & \textbf{99.06} & \textbf{99.06} \\
                                                      & Masking-alone     &  \textbf{81.56} & \textbf{96.83} & \textbf{96.93} & \textbf{96.91} & \textbf{34.78} & \textbf{95.75} & \textbf{99.66} & \textbf{99.72} & \textbf{84.74} & \textbf{87.68} & \textbf{87.79} & \textbf{87.79} & \textbf{81.86} & \textbf{99.05} & \textbf{99.06} & \textbf{99.06} \\

\hline

\end{tabular}
\vspace{-2mm}
\end{table*}

\textbf{Metrics.} We use three metrics for evaluation in this section. 
\begin{enumerate}[leftmargin=*]
\itemsep0em 
  \item \emph{Robust Accuracy} is the prediction accuracy on all the AEs.
  \item \emph{Mitigation FPR} is the (reduced) FPR from the two-staged combination of detection and mitigation (explained in Section~\ref{sec:reduce-fp}).  
  \item \emph{Mitigation success recall} is the detection recall from the combination of detection and mitigation (Section~\ref{sec:reduce-fp}) - we distinguish this from the  \emph{detection success recall}, which is the detection recall from the detection technique alone.
  \emph{Mitigation success recall} gives the final amount of adversarial examples detected.
\end{enumerate}

\textbf{Result.} 
{
Table~\ref{tab:mitigation} shows \sysname's mitigation performance on all datasets. The results are averaged across patches of different sizes (5\% to 7\%). 
The detection performance is higher on larger patches as these patches have higher attack success rates  (difference between the largest and smallest patch is about 9\%). The mitigation performance is consistent across different patch sizes (differences 
less than 2\%). 
We consider two mitigation strategies, (1) masking-alone and (2) GAN-based recovery (which first performs masking and then use GAN to recover the contents from the mask) (\sysname).  
}

\emph{Robust accuracy}: Masking with GAN-based restoration is able to yield higher robust accuracy than masking alone. This is because GAN-based restoration synthesizes the missing semantic contents in the mask for the network to make a correct prediction. 
The only exception is when 25\% or 50\% of the pixels are masked, where masking alone has higher robust accuracy than GAN-based recovery.  
This is because the GAN relies on the regions outside the mask as the context to synthesize the contents. When the masking percentage is small, a large portion of the  adversarial pixels remain intact, 
and thus the GAN cannot reconstruct the contents correctly. 
In this case, it is better to mask the perturbations to shield their contributions to the prediction,  rather than GAN-based recovery as done by \sysname.

{We also notice that the robust accuracy by \sysname on CelebA is lower than that on the other  datasets, which is because the GAN needs to synthesize the correct facial features belonging to a particular celebrity's face to enable correct identity prediction. 
This is a much more challenging task for the GAN than for the other three datasets, and hence \sysname yields a lower robust accuracy. 
\sysname achieves the highest robust accuracy on ImageNette, as it is a 10-class dataset, and performing correct image classification on this dataset is easier than on the other complicated datasets such as the 1000-class ImageNet.
}

\emph{Mitigation FPR}: 
GAN-based recovery achieves low FPR, because the restored inputs are more similar to the original benign inputs than the masked inputs (in the latter case many features are simply masked). Therefore predictions on the original and restored inputs are more likely to be the same, which is not the case for inputs that are merely masked. 
{We also see that \sysname consistently achieves very low FPRs on all the datasets.}

\emph{Mitigation success recall}: While masking alone is able to achieve higher detection recall compared to GAN-based recovery when the masking percentage is small, the difference becomes negligible when the masking percentage increases. 
This is because when the masking percentage is low, the masked images are more likely to have a label \emph{different} from that of the original image; while the restored images are more likely to have the \emph{same} label as the original image - this is similar to the reason why robust accuracy from masking alone is higher than that from GAN-based recovery for 25\% masking. 
However, when the masking percentage increases, both the masked and restored images are likely to have labels that are different from that of the original image - thus the difference becomes negligible between both approaches.
{We see that \sysname is highly effective in detecting adversarial examples on all the datasets.}

\begin{mdframed}[linewidth=2pt]
{
The GAN-based recovery strategy  
outperforms the masking-alone strategy with higher robust accuracy and lower FPRs.
}
\end{mdframed}

\emph{Trade-off by varying masking percentage}:
Our results also show that the proposed parametric masking is able to moderate the balance between different metrics, based on which the defender can adjust \sysname to prioritize different outcomes.
For instance, if the defender's goal is to detect/mitigate adversarial attack while \emph{minimizing} FPR on the benign inputs, he/she can perform recovery on 50\% of the suspicious features, which is able to detect over 91\% of the adversarial examples, achieve a robust accuracy of over 70\% with a FPR of less than 0.5\% (all on average). 
On the other hand, the defender who wants to \emph{maximize} \sysname's performance can perform recovery on 100\% of the suspicious features, which on average yields the highest robust accuracy (79.73\%) and detection success recall (95.86\%) with a slightly higher FPR (0.71\%).  

\begin{mdframed}[linewidth=2pt]
 \sysname is able to balance between different performance metrics, by varying the percentages of the GAN-based recovery. 
\end{mdframed}

\subsection{RQ3 - Comparison with Related Techniques }
\label{sec:comparison}
We consider four related defenses against patch attacks below (and compare with two more trajan-attack defenses in Appendix~\ref{sec:trojan-defense-comp}).

\textbf{1. Localized Gradient Smoothing~\cite{naseer2019local}.}
Naseer et al. propose local gradient smoothing (LGS) to neutralize the effect of adversarial patch pixels.
They first perform normalization over the gradient values, and then use a moving window to identify high-density regions (based on certain thresholds), which will be smoothed out to suppress the influence of the adversarial pixels.
We follow \cite{naseer2019local} to set the threshold as 0.1 and smoothing factor as 2.3.

\textbf{2. Adversarial training.}
Adversarial training (AT) increases the robustness of DNNs by explicitly training the networks to be robust against the patch attack~\cite{rao2020adversarial,wu2020defending}. 
We adopt the approach from prior work  ~\cite{rao2020adversarial,wu2020defending} to conduct AT on ImageNette.
We first train the models on clean images, which is then used for adversarial training. 
For each DNN, we train three different models, one for each patch size.  
We train the models by using the SGD optimizer and varying different hyperparameters such as learning rate, momentum, dropout, number of epochs, batch size.

\textbf{3. SentiNet~\cite{chou2020sentinet}.}
Chou et al. propose SentiNet for detecting patch attacks.
Sentinet first uses a selective search image segmentation to generate a list of class proposals, i.e., input segments corresponding to different classes.
It then extracts the salient maps from the class proposals and identifies the unique features, by subtracting the common regions in the saliency maps, which is then overlaid to a set of new images. 
It then replaces the extracted features with inert patterns such as Gaussian noise in order to distinguish between adversarial and benign features.
The final attack detection is based on statistical analysis on two metrics: 
(1) the number of misclassified images from images with the unique extracted features, and (2) the average confidence values from images with inert patterns. 
We follow \cite{chou2020sentinet} to overlay the salient features from the test image to 100 new images, which is to calculate the statistics for detecting adversarial examples. 
We randomly sample 400 clean images to compute the detection threshold for detection. 
{
SentiNet~\cite{chou2020sentinet} is the most closely related technique to \sysname, and we compare both techniques in detail below.
}

{
\textbf{4. PatchGuard~\cite{xiang2020patchguard}.}
Xiang et al.~\cite{xiang2020patchguard} propose PatchGuard, a certified defense technique against patch attacks. The main idea is to enforce small receptive field in the DNNs, and secure feature aggregation by masking out the regions with the highest sum of class evidence (as these regions are more likely to be manipulated by adversarial patch to dominate the prediction).  
}

\begin{table}[t]
\begin{threeparttable}
\caption{Comparison with LGS~\cite{naseer2019local}, SentiNet~\cite{chou2020sentinet} and PatchGuard~\cite{xiang2020patchguard}.
Better results are highlighted in \textbf{bold}.}

\label{tab:comparison-1}
\centering  
\footnotesize

\begin{tabular}{|c| cccc |}
\hline
\multirow{2}{12em}{\centering \textbf{Metric} (\%)}  & \multicolumn{4}{|c|}{\textbf{Technique}}  \\

%\cline{3-10}
  &  LGS  & SentiNet & PatchGuard & \sysname  \\
\hline

\textbf{Detection Recall}   & N/A     & 73.04 & N/A   & \textbf{96.89}  \\
\textbf{Robust Accuracy}    & 53.86   & N/A   & 11.70 & \textbf{77.47}  \\
\textbf{False Positive}             & 12.14   & 7.66  & 44.67 & \textbf{1.74}  \\

\hline
\end{tabular} 

\end{threeparttable}
\vspace{-4mm}
\end{table}

\textbf{Result.} 
We compare \sysname (GAN-based recovery with 100\% masking) with LGS, SentiNet and PatchGuard on ImageNet. 
Table~\ref{tab:comparison-1} shows the average results for patches of different sizes. 

\textbf{1. Comparison with LGS. }
LGS achieves an average robust accuracy of 53.86\%, which is considerably lower than that of 77.47\% by \sysname. This is because: (1) not all adversarial patch regions would stand out as the high-density region after normalization by LGS; and (2) LGS uses gradient smoothing as the mitigation strategy, which is inferior to GAN-based recovery by \sysname that can reconstruct the semantic contents from the corrupted regions. 
LGS also incurs a very high FP of 12.14\%, which is because the natural features in the benign examples may also be identified as high-density regions and hence LGS incorrectly perform gradient smoothing on these regions. In contrast, \sysname incurs an FP rate of only 1.74\%. 

\textbf{2. Comparison with SentiNet. }
SentiNet detects 73\% adversarial examples while \sysname detects 96.89\%, which yields an improvement of 32.7\%. 
Further, SentiNet has an FPR of 7.66\% while \sysname has only 1.74\% (a 77.3\% reduction). 
Hence, \sysname outperforms SentiNet by {\em having a higher detection rate on adversarial examples, and achieving a much lower FPR.} 
{We next qualitatively compare both techniques to understand their significant performance difference.
}

The low detection rate of SentiNet is due to its poor identification of the adversarial patch region. 
Specifically, SentiNet extracts the adversarial patch region by subtracting the common regions of the saliency maps belonging to different classes. 
However, the adversarial patch may reside in the common regions of different saliency maps, and thus the patch region will be removed after subtraction, thereby remaining \emph{undetected}.

Though \sysname also uses saliency map in its detection, \sysname follows a {different} principle to locate adversarial patch from the saliency map and proposes a robust suspicious feature detection method (e.g., a pre-processing technique to highlight adversarial patch region) that can reliably locate the adversarial patch \emph{and} verify its maliciousness through a guided feature transplantation and prediction comparison. This allows \sysname to detect substantially more AEs (32.7\% more).

The high FPR in SentiNet is because SentiNet overlays the suspicious features to a \emph{random} region of an image, which could cause FPs when the salient features occlude the image's natural features. 

Instead, \sysname achieves low FPR through: 1) strategically transplanting the salient features to the \emph{least-salient} feature region of the image; and 2) using the generative power of GAN to reduce FPR. 
Both innovations combined allows \sysname to achieve a much lower FPR than SentiNet (77.3\% lower).

\textbf{3. Comparison with PatchGuard. }
PatchGuard provides provable robust accuracy but has a robust accuracy of 11.7\% and an FPR of 44.67\%. In contrast, \sysname has a  77.47\% (empirical) robust accuracy with only 1.74\% FP. This is because PatchGuard provides a (provable) lower bound of the adversarial robustness and the high FPR is due to the small receptive field enforced by PatchGuard, which causes considerable clean accuracy drop as in \cite{xiang2020patchguard}.

\textbf{4. Comparison with adversarial training (AT).}   
{
AT requires training for \emph{each} target class, which is challenging as attackers may target diverse classes.
In contrast, \sysname does not require any training, and is agnostic to the target classes of the attack.}
Therefore, we compare the performance of AT and \sysname when the attacker targets different classes, by training multiple 7\% patches targeting \emph{different} labels on ImageNette.

{
Table~\ref{tab:comparison-3} shows that AT's performance degrades as the number of target classes increases, on both robust accuracy and FP.
This is because with more target classes, the learning objective for AT becomes increasingly difficult - this is similar to how common DNNs would yield lower accuracy on a 1000-class dataset than on a simple 10-class dataset. 
On the other hand, we see that \sysname achieves consistently high performance in terms of both robust accuracy and FP across attacks targeting different classes.
Further, \sysname yields significantly better performance than AT in all cases.
}

\begin{table}[t] 
\caption{Comparison with AT in defending against \emph{multiple} target classes.
Better results are highlighted in \textbf{bold}.}
\label{tab:comparison-3}
\centering  
\footnotesize
\begin{tabular}{|l| ccc |  ccc |}
\hline
\multirow{2}{3em}{\textbf{Metric} (\%)}  & \multicolumn{3}{|c|}{\textbf{Adversarial training}} & \multicolumn{3}{|c|}{\textbf{\sysname}}  \\
  &  1 target & 3 targets  & 5 targets &  1 target & 3 targets  & 5 targets   \\
\hline
\textbf{Robust Acy.}                & 92.29   & 84.49    & 78.97 & \textbf{95.34}  & \textbf{94.18}   & \textbf{94.15}  \\
\textbf{False Positive}             & 20.23   & 24.22    & 27.05 & \textbf{0.72}   & \textbf{0.87}    & \textbf{0.88}  \\

\hline
\end{tabular} 
\vspace{-2mm}
\end{table}

\subsection{RQ4 - Physical-world Patch Attacks }
\label{sec:physical-attack}
In this RQ, we evaluate the effectiveness of \sysname against physical-world patch attacks.
We use the printable adversarial patch from \cite{brown2017adversarial}. We printed it out,   placed it next to the cell phone object (a iPhone 6s device) at various locations, and captured its video. 

Fig.~\ref{fig:physical} shows the video frames in our evaluation. 
Both videos with and without patches contain around 430 frames. 80\% of the frames with patches successfully caused the targeted misclassification.  
We increase the length of the detection box to 142 as the physical patch occupies more pixels in the images than the digital patch~\cite{brown2017adversarial}.

\begin{figure}[t]
\centering
  \includegraphics[ height=0.7in, width=3.0in]{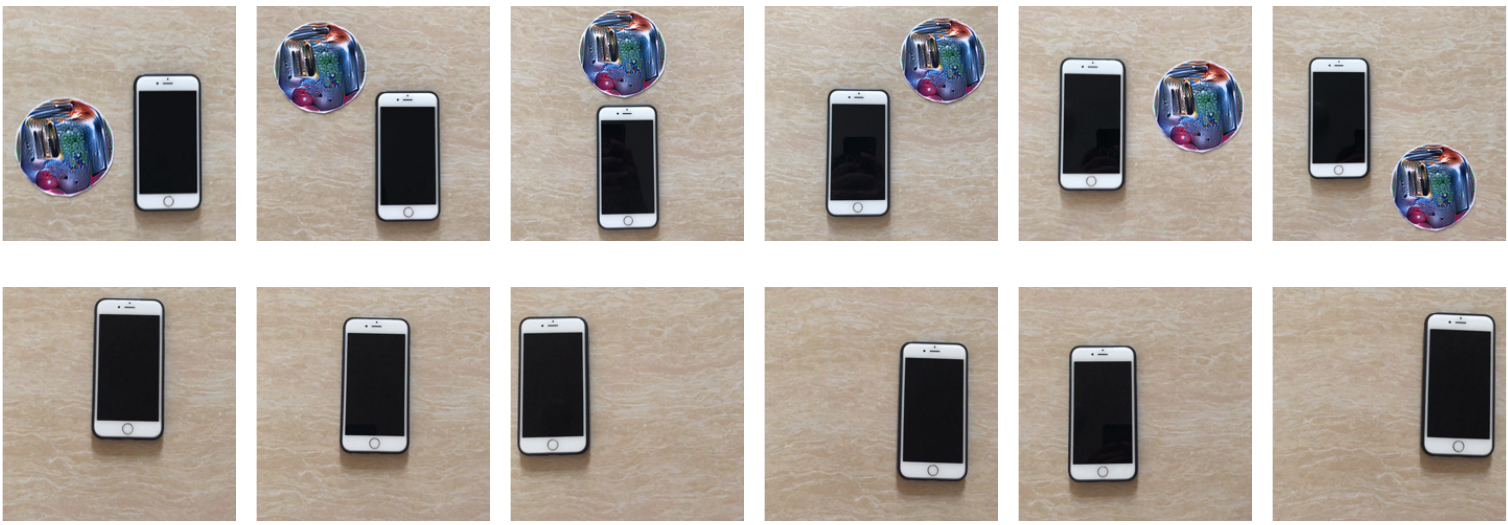}
\caption{Video frames for a cell phone object with (top row) and without (bottom row) a physical adversarial patch. 
}
\label{fig:physical}
\vspace{-2mm}
\end{figure}

As before, we evaluate the effectiveness of \sysname  in terms of robust accuracy, mitigation success recall and mitigation FPR.
The results are shown in Table~\ref{tab:physical-res}. 

\emph{Robust accuracy.} Unlike the previous evaluation on digital patch, we can see from Table~\ref{tab:physical-res} that a low masking percentage is able to yield a high robust accuracy for the physical attack (this is low for the evaluation on digital patch). 
This is because the perturbations in the physical patch are more susceptible to masking and recovery compared with the digital patch that is directly applied to the image.    
Perturbations in the physical patch need to undergo camera transformations, which makes the perturbations more amenable to being mitigated by \sysname.
Thus, even a low masking percentage in \sysname is able to effectively mitigate the physical patch attack.

In addition, the robust accuracy from 75\% and 100\% masking is lower than that from 50\%, which is unlike the trend in the previous evaluation for digital patches.
This is because the detection box is larger, and hence a higher masking percentage means more contents are masked for the GAN to recover, which leads to degraded quality of the recovered images and thus the DNN is unable to infer the correct label. 
Therefore, 50\% masking yields the highest robust accuracy of 95.32\%.

\emph{Mitigation FPR} yielded by \sysname ranges from 1.15\% to 5.52\%. Similar to the digital patch, the FPR is higher when the masking percentage is higher. 

\emph{Mitigation success recall} yielded by \sysname ranges  
from 86.84\% to 95.91\%. The trend is similar to that of digital patch, i.e., the success recall is low when the masking percentage is low (and vice versa).

\begin{table}[t]
\caption{\sysname's performance on \emph{physical patch attack}.}
\label{tab:physical-res}
\centering 
%\normalsize
\footnotesize
\begin{tabular}{|c |cccc |}
\hline
\multirow{2}{*}{\textbf{Metric} (\%)} & \multicolumn{4}{|c|}{\textbf{Masking percentage}}  \\
 
  &  25\% & \textbf{50\%} & 75\% & 100\%  \\
\hline

 {\textbf{Robust Accuracy}}  &  82.46 & \textbf{95.32} & 93.86 & 81.87    \\ 

\hline
 {\textbf{Mitigation FPR} }   &  1.15 & 2.99 & 3.45 & 5.52     \\ 
 
\hline
 {\textbf{Mitigation Success Recall}}   &  86.84 & \textbf{95.91} & 95.91 & 95.91     \\ 
                                                                        
\hline

\end{tabular}
\vspace{-2mm}
\end{table}

\begin{table}[t]
\caption{\sysname's performance on \emph{attacks targeting 5 labels}. }
\label{tab:defend-5-class}
\centering 
%\normalsize
\footnotesize
\begin{tabular}{|c| c | c| c | c  |}
\hline

{\textbf{Metric} (\%)}  &  {\textbf{ImageNet}} & {\textbf{ImageNette}} & {\textbf{CelebA}} & {\textbf{Place365}}  \\        

\hline 

 {\textbf{Robust Accuracy}}   & 79.27 & 94.15 & 71.69 & 77.32     \\ 

\hline 
 {\textbf{Mitigation FPR}}   &  1.61 & 0.88 & 0.00  & 0.20  \\ 

\hline 

{\textbf{Mitigation Succ. Recall}} & 99.54 & 98.57 & 96.25 & 93.98 \\                                                                  

\hline

\end{tabular}
\vspace{-2mm}
\end{table}

\subsection{{RQ5 - Attacks Targeting Different Labels} }
\label{sec:atk-diff-label}
This section evaluates \sysname against attacks that target different class labels. 
For each target label, we need to perform training to generate the universal adversarial patches. Training  is  highly time-consuming, and hence, for each dataset, we train five 7\% patches targeting different labels. 
Note that training is only needed for creating the adversarial patches, and not for \sysname.

{
Table~\ref{tab:defend-5-class} shows the results.  \sysname consistently achieves high performance in detecting and mitigating patch attacks targeting different labels, and with very low FPR. 
On average, \sysname detects over 97\% of the adversarial examples, achieves a robust accuracy of over 80\%, with only 0.67\% FPR.
This high accuracy is because \sysname works by comparing the prediction label before and after feature transplantation, which is \emph{agnostic} to the exact target label. 
}

\subsection{RQ6 - Adaptive Attacks}
\label{sec:adaptive-attack}
{
We now evaluate two adaptive adversaries attempting to fool the DNNs even under \sysname by: 
(1) evading the detection by manipulating the saliency map; (2) evading the mitigation by generating strong perturbations that can continue to fool the DNN.
}

\subsubsection{Evading the Detection}
\label{sec:adaptive-evade-detection}
{
\sysname  uses the saliency map to detect adversarial patches, and the saliency map can be manipulated by an  adversary as shown by prior work~\cite{zhang2020interpretable,ghorbani2019interpretation}. 
{However, they  
consider adversarial perturbations over the entire image, while we consider \emph{localized} perturbations. Therefore, we adapt their approach~\cite{zhang2020interpretable,ghorbani2019interpretation} to manipulate the localized saliency map so that the adversarial patch \emph{will not} be identified as the most salient features (by reducing its influence on the output), hence evading \sysname. 
}  
}

\textbf{Attack setup.}
The influence on the output is derived from the saliency map.  
This attack can be formulated as the following objective function during the generation of the adversarial patch:

\begin{equation}
\label{eq:evasion-gen}
\delta = \argmax_{\delta}\mathbb{E}_{x \sim X, l \sim L}(\text{log}\mathrm{Pr}(y=y^{adv}|x') - \beta\Arrowvert\hat{M}^*_j(x) - m^*_0 \Arrowvert^2_2  ),
\end{equation} 

where $\hat{M}^*_j(x)$ is defined as the saliency map on the region where the adversarial patch resides (not the entire saliency map), and $m^*_0$ is a mask in the same size of the adversarial patch and filled with 0, $\beta$ is a hyperparameter to balance different loss terms. 
The first term is to cause targeted misclassification, while the second term's goal is to let $\hat{M}^*_j(x)$ have \emph{small} influence to evade detection by forcing the values within $\hat{M}^*_j(x)$ to be close to 0.  
The adversary stops the optimization once the patch succeeds in evading the detection box.

The second term can be viewed as manipulating the Hessian matrix of $F(x)$, whose values are all zero for DNNs with ReLu activation functions~\cite{zhang2020interpretable,ghorbani2019interpretation}. 
Therefore, we replace the ReLu function with a parametric softplus function when calculating the gradients~\cite{ghorbani2019interpretation,xie2020smooth}:  
$f(x) = \frac{1}{\alpha} \text{log}(1+\text{exp}(\alpha x))$, where  
$\alpha$ is the hyper-parameter to control the shape of the curve and is set as 10 (following \cite{xie2020smooth}). 
Finally, we only use the parametric softplus for backward propagation, and use ReLU for the normal forward pass.

To be conservative, we consider the 7\% patch, which allows the attacker to inject larger  perturbations to evade \sysname.  
We choose 200 samples for training the adversarial patch, 500 steps per sample and 20 epochs in total.
For each dataset, we choose $\beta \in [0.1, 0.5, 1, 5]$ and choose the one yielding the highest attack success rate. 

Equation~\ref{eq:evasion-gen} requires several forward and backward passes for calculating the saliency map $\hat{M}^*_j(x)$, which is much more time-consuming than the original optimization  (Equation~\ref{eq:attack-gen}).
Therefore, we reduce the sampling size $n$ in Equation~\ref{eq:smoothgrad} from 50 to 5 for faster training.  
We experimentally verified that the smaller sampling size $n$  does not significantly affect the resulting saliency map, and that we can still find all the salient features.   
Under this setting, it took around 18 days to generate an adversarial patch on Place365 dataset, compared to about 540 days with our previous setup (estimated).

\textbf{Result.}
We compare the attack success rate of the patches generated from the undefended models and the ones guarded by \sysname in Fig.~\ref{fig:adp-evade-detection}.
When \sysname is used, the adaptive attacker who attempts to evade \sysname's detection suffers a significant drop in attack success rate, from 99\% to just 4.9\% (on average). 
This is because in Equation~\ref{eq:evasion-gen},  
the first term aims to \emph{increase} the influence on the final prediction to manipulate the output label; 
while the second term \emph{reduces} the influence on the output. This equation constrains the adaptive attacker, who cannot evade detection without also  significantly degrading the attack's effectiveness in the process.

\begin{figure}[t]
\centering
  \includegraphics[ height=0.8in,trim=3 3 3 3,clip]{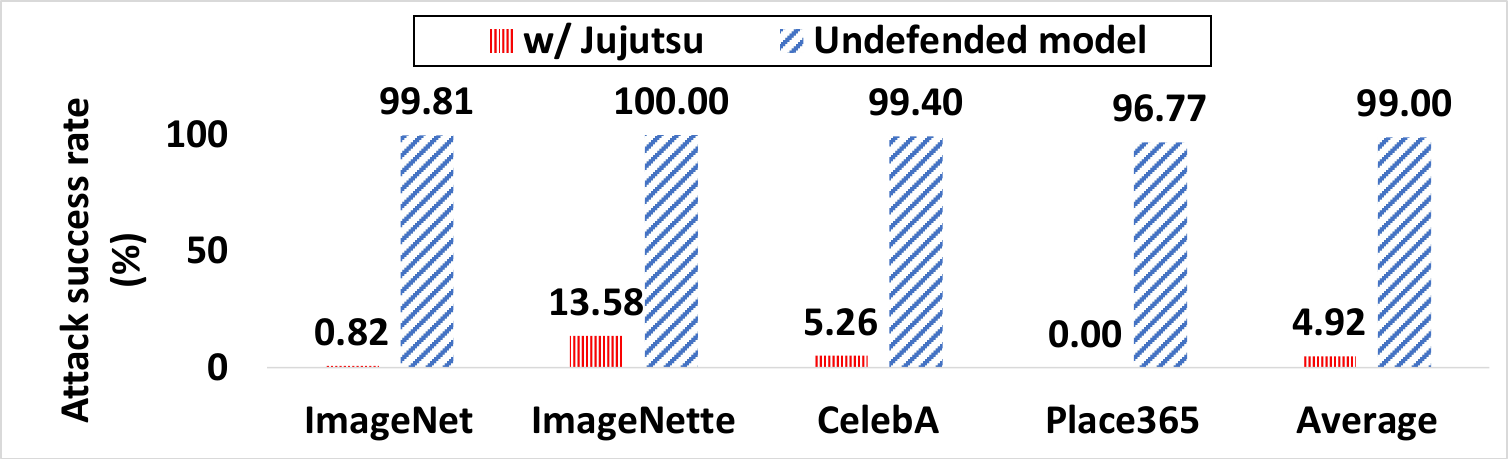}
\caption{Attack success rate (ASR) for adaptive attack to evade \sysname's \emph{detection}. The lower the better. }
\label{fig:adp-evade-detection}
\vspace{-4mm}
\end{figure}

\subsubsection{Evading the Mitigation}
\label{sec:adaptive-evade-mitigation}
We now consider a second adaptive attacker who attempts to cause targeted misclassification even if the adversarial examples are detected.  
Because our masking strategy is parametric, the adaptive attack would be unsuccessful if we perform recovery on the entire set of suspicious features since all of the adversarial perturbations would be removed. 
Therefore, we study whether the adaptive attack could succeed if the defender masks only 50\% or 75\% of the suspicious features.

\textbf{Attack setup.}
The adversary's goal is to generate an adversarial patch that can survive under partial masking (50\% or 75\% masking). 
To model the masking of $x\%$ of the suspicious features, we randomly set $x\%$ of the values that are non-zero within the mask $m \in \{0,1\}^n$ to be 0, so that those positions marked with a 0 will not be available for manipulation by the attacker. Therefore, the attacker can use only the remaining perturbations  
to cause misclassification.

Similar to Section~\ref{sec:adaptive-evade-detection}, we consider the 7\% patch to maximize the attack's influence.  
For the masking percentage of 50\%, we use 2000 images, a maximal step of 1000 and 30 epochs in total, which is in line with our standard attack generation as in Section~\ref{sec:attack-setup}. This is because evading the mitigation does not require several forward and backward passes for each step as in Section~\ref{sec:adaptive-evade-detection}, and thus we can use more images and optimization steps as well as epochs. % to generate the perturbations.

For the 75\% masking percentage, we use a maximal epoch of 20, because our evaluation shows that the training is not able to make any progress, and the attack success rate is consistently close to 0\%.

\textbf{Result.}
{
Fig.~\ref{fig:adp-evade-mitigation} shows the percentage of adversarial examples that succeed in causing misclassification despite \sysname. 
We can see that the ASRs degrades under the adaptive attack. 
}

{
When 75\% of the perturbations are masked, it is almost infeasible for the attacker to generate a  successful adversarial patch, and hence the success rate is near 0\% on all datasets. 
}

{
When 50\% of the perturbations are masked, the ASRs are however much higher, ranging from 28\% to around 60\%. 
This is because many of the detected adversarial examples will be mis-identified as benign after GAN-based recovery with 50\% masking is performed. 
When the masking percentage is low, the adversarial perturbations will remain intact and they can continue to cause misclassification on the restored images - \sysname will mis-identify them as benign. 
}

{
Nevertheless, \emph{the defender can further thwart the attack by increasing the masking percentage to 75\% or 100\% in \sysname}, which only comes at a cost of slightly higher FPRs (0.23\% higher) and is able to mitigate most of the adversarial examples. For example, by performing GAN-based recovery on 100\% of the suspicious features, the average attack success rate is reduced from 48\% to 22\%, which is significantly lower than that on the undefended models (99\%). 
}
 
Our results show that \sysname's performance can be degraded by the adaptive adversary via {deliberately} reducing the success rate of patch attacks.
For instance, the adaptive patch generated on ImageNet for 50\% masking has a success rate of 76.8\%, which is lower than the 99\% of the non-adaptive patch. 
This results in 23.5\% of adversarial examples going  undetected by \sysname. 
However, we note that doing so would significantly \emph{constrain} the adversary's ability in fooling the DNN, and hence make the DNN  significantly less susceptible to patch attacks. 
On average, the attack success rate is reduced from 99\% to 0.73\%$\sim$ 22\% across the four datasets.

\begin{figure}[t]
\centering
  \includegraphics[ height=0.8in,trim=3 3 3 3,clip]{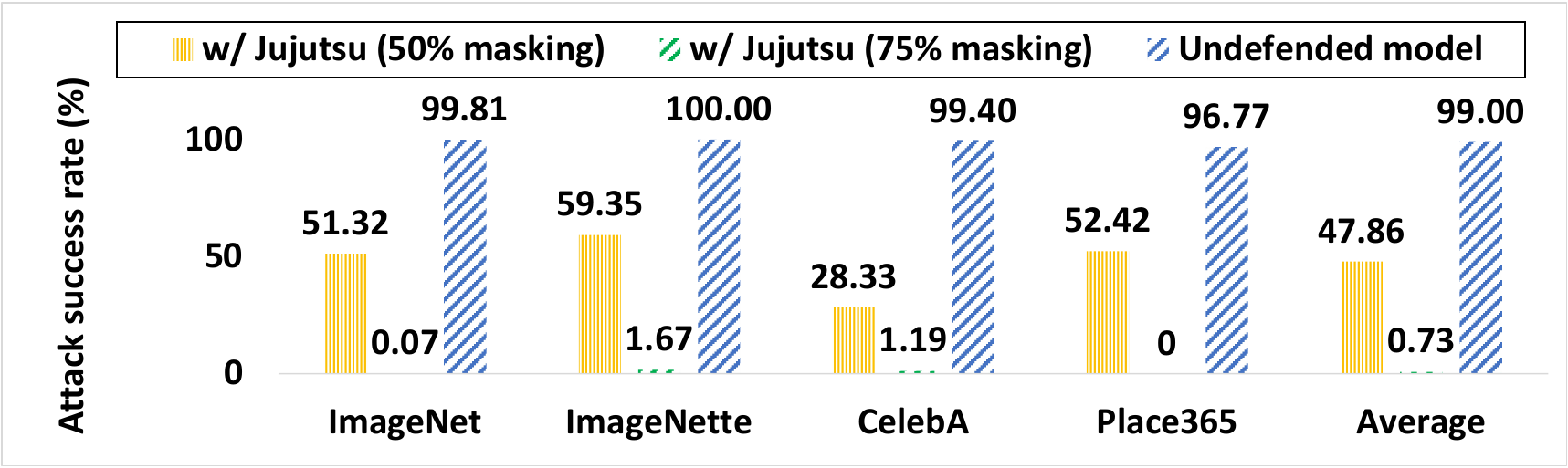}
\caption{Attack success rate (ASR) for adaptive attack to evade \sysname's \emph{mitigation}. The lower the better. }
\label{fig:adp-evade-mitigation}
\vspace{-4mm}
\end{figure}

\section{Discussion}

\label{sec:discussion}

This section first discusses the limitation of \sysname, followed by the evaluation of \sysname's performance on more DNN models and more different patch sizes. We present an ablation study in Appendix~\ref{sec:ablation-study}.

\subsection{{Limitation}}
\label{sec:limitation}
First, \sysname incurs overhead in its attack detection and mitigation, and we report its overhead in Appendix~\ref{sec:overhead} due to space limitations.

{
Second, \sysname employs PICnet~\cite{zheng2019pluralistic} in its attack mitigation, whose performance also affects \sysname's mitigation performance. 
For example, \sysname has lower robust accuracy on the CelebA dataset (than other dataset) as it is challenging to synthesize specific human faces with PICNet for the DNN to make correct predictions. 
Nevertheless, as image inpainting with GAN is an active research area, we believe this issue can be further alleviated by incorporating recent research results. 
For instance, Li et al.~\cite{li2022mat} recently introduced a transformer-based model that yields superior inpainting performance on diverse tasks, which may be leveraged by \sysname to enhance its mitigation performance. 
}

Finally, there are other attack variants outside our threat model. We discuss next  how \sysname may be extended to handle them. 

\emph{Multiple patches.}
We focus on defense against single-patch attacks, on which existing defenses~\cite{naseer2019local,rao2020adversarial,wu2020defending,chou2020sentinet,xiang2020patchguard} have very limited success. 
However, multi-patch attacks are also possible and one potential solution to handle them is to \emph{iteratively} perform detection and mitigation until only the benign features are left in the images.
This can be achieved by checking whether the current suspicious features \emph{fail} to cause misclassification on the hold-out inputs, which occurs when the suspicious features are benign.  

{To validate this, we evaluate the above extension of \sysname against 2-patch and 3-patch attacks and it is still able to achieve good performance - see Appendix~\ref{sec:multi-patch}}.  

\emph{Patches in different shapes.} 
{
Similar to prior work~\cite{naseer2019local,xiang2020patchguard,levine2020randomized,mccoyd2020minority}, we assume a square or circle patch, and conduct extensive evaluation on these attacks. 
But patches in other shapes such as rectangular one are also possible, which is a limitation of \sysname (and also of other defenses~\cite{naseer2019local,xiang2020patchguard,levine2020randomized,mccoyd2020minority}). 

Nevertheless, we evaluate how \sysname can be extended to defend against rectangular patch (by assuming a rectangular detection box) - see Appendix~\ref{sec:diff-patch-shape} for details.   
We leave the generalization of \sysname to different patch shapes to future work (e.g., instead of using one single detection bounding box, can we use multiple detection boxes in different shapes to cover different potential patches and flag an attack if any of the suspicious features extracted from different boxes is deemed as adversarial?).
}

\emph{Untargeted attacks.}
We focus on targeted attacks, which allow the adversary to manipulate the DNNs in a controlled manner. Other work ~\cite{athalye2018obfuscated,kannan2018adversarial}  has recommended evaluating targeted attacks for large-scale datasets like ImageNet, as untargeted attacks may cause misclassification of very similar classes (e.g., images of two very similar dog breeds). 

We tried extending \sysname to defend against untargeted attack by directly performing attack mitigation on the suspicious features (without the guided feature transplantation and prediction comparison for attack detection), and use the prediction label on the recovered image as the final output, but we had very limited success (see Appendix~\ref{sec:untargeted} for details). 
Therefore, future work may combine \sysname with other defenses (e.g., using pre-defined thresholds~\cite{naseer2019local} or small masking on the test image~\cite{xiang2021patchcleanser}) to first characterize benign and adversarial examples, and then use \sysname's mitigation technique to perform attack recovery.

\begin{table}[htb]
\caption{\sysname's performance on 6 different DNNs. }
\label{tab:diff-networks}
\centering 
%\normalsize
\footnotesize
\begin{tabular}{|c| c | c| c | c  | c | c |}
\hline

\multirow{2}{4em}{\textbf{Metric (\%)} }  &  \multirow{2}{3em}{\textbf{Wide-ResNet}} & \multirow{2}{3em}{\textbf{Dense-Net121}} & \multirow{2}{3.5em}{\textbf{Squeeze-Net}} & \multirow{2}{2.5em}{\textbf{VGG16}} & \multirow{2}{3em}{\textbf{ResNet-152}} & \multirow{2}{3em}{\textbf{Google-Net}}    \\        
            & & & & & & \\
\hline
 {\textbf{Clean Accuracy}}   &  78.27 & 71.58 & 54.74 & 72.03 & 76.84 & 67.17   \\ 

\hline 

 {\textbf{Robust Accuracy}}   &   82.21 & 76.6 & 69.03 & 78.33 & 79.33  &  73.44  \\ 

\hline 
 {\textbf{Mitigation FPR}}   &   2.13 & 2.39 & 0.67 & 1.91 & 2.38 &  3.47   \\ 

\hline 

{\textbf{Mitigation}} &  \multirow{2}{3em}{99.92} & \multirow{2}{3em}{98.11} & \multirow{2}{3.5em}{98.43} & \multirow{2}{2.5em}{99.05} & \multirow{2}{3em}{97.75}  &  \multirow{2}{3em}{99.92}  \\ 
     \textbf{Succ. Recall}       & & & & & & \\
\hline

\end{tabular}
\vspace{-2mm}
\end{table}

\subsection{{Evaluation on More DNN Models}}
\label{sec:more-dnn}
{
This section evaluates \sysname's performance on 6 more DNN models (Wide-ResNet, DenseNet121, SqueezeNet VGG16, ResNet152 and GoogleNet) on ImageNet (all are the pre-trained models from the TorchVision library). 
We report the results in Table~\ref{tab:diff-networks}. 

\sysname consistently achieves high detection performance (average detecting over 98\% AEs), low FPRs (average 2.16\%), and high robust accuracy (average 76.49\%). 
The robust accuracy varies between different models because it is also related to the prediction performance of the model, e.g., SqueezeNet has a relatively low clean accuracy, which also leads to lower robust accuracy compared with other models. 
Our results show that \sysname's outstanding defense performance can generalize across different DNN models.
}

\subsection{{Evaluation on Larger Patches}}
\label{sec:larger-patch-eval}
{
Section~\ref{sec:evaluation} evaluates 5\%$\sim$7\% patches. In this section we evaluate \sysname against larger patch: 8\%, 9\% and 10\% 
We report the average results as there is no major variation between different patch sizes.
 
On average, \sysname detects 99\% of the adversarial examples, achieve a robust accuracy of 81.77\% with a FPR of only 0.71\%. 
\sysname yields slightly better performance on these larger patches (compared with the results in Table~\ref{tab:mitigation}) because these larger patches are able to achieve slightly higher attack success rate. 
Therefore, \sysname's high detection and mitigation performance and low FPs can generalize to larger adversarial patches.
}

\section{Related Work}
\label{sec:related-work}

Defenses against adversarial attack can be  divided into certified~\cite{chiang2020certified,levine2020randomized,xiang2020patchguard,mccoyd2020minority,xiang2021patchcleanser} and empirical  
defenses~\cite{naseer2019local,hayes2018visible,rao2020adversarial,xu2020lance,wu2020defending,jha2019attribution,chou2020sentinet}.  

\emph{Certified defenses. } Chiang et al.~\cite{chiang2020certified} propose the first certified defense against patch attack by Interval Bound Propagation, which constrains the influence of the adversarial pixels in the hidden layers to compute a lower bound of the robustness. Levine et al.~\cite{levine2020randomized} introduce de-randomized smoothing (DS) to build a smoothed classifier whose prediction is based on the ensemble of local prediction on pixel patches. 
Xiang et al.~\cite{xiang2020patchguard} propose PatchGuard which is based on enforcing small receptive field and robust feature masking. PatchCleanser~\cite{xiang2021patchcleanser} is a provable defense whose main idea is that clean examples are more robust than AEs under small masking and the AEs can be mitigated via a two-round masking. 

\emph{Empirical defenses. }
Naseer et al.~\cite{naseer2019local} propose local gradient smoothing (LGS) to mitigate patch attacks by identifying the regions with high gradient magnitude, and smoothing out regions whose gradient values are greater than a certain threshold.  
Hayes et al.~\cite{hayes2018visible} introduce digital watermark (DW), which uses pre-defined thresholds to scan the saliency map to detect and remove important pixels (hence the adversarial pixels are also removed).  
Jha et al.~\cite{jha2019attribution} detect patch attacks by selectively masking the top-$k$\% salient features and comparing the prediction labels. 
While useful in detecting adversarial examples, these techniques also incorrectly flag many benign samples as being adversarial because the natural features in benign samples may also exceed the pre-defined thresholds or remain as the top-$k$\% salient features. Hence these techniques suffer from high  FPRs. 
Chou et al.~\cite{chou2020sentinet} propose SentiNet for detecting patch attacks based on model interpretability and statistical analysis, which provides limited detection performance and does not support attack mitigation.  
Adversarial training, a standard adversarial defense technique, can also be used to defend against patch attacks~\cite{rao2020adversarial,wu2020defending}. 
But it also suffers from high FPRs and performance degradation when the adversary targets multiple labels.  
Unlike prior work, \sysname is a two-staged defense that provides both attack detection and mitigation, and yields superior detection and mitigation performance with very low FPRs.

\section{Conclusion}

This work proposes \sysname, a technique to detect and mitigate robust and universal adversarial patch attacks against image classification DNNs.
\sysname  accurately locates adversarial patch and distinguishes it from benign samplse, and uses generative adversarial networks to reconstruct the ``clean'' examples from adversarial examples.
Our extensive evaluation on four datasets and comparison with four defenses show that  \sysname achieves superior detection and mitigation performance, with low false positives. \sysname also defends against physical-world and adaptive attacks.

\section*{Acknowledgements}
This work was funded in part by the Natural Sciences and Engineering Research Council of Canada (NSERC), and a Four Year Fellowship from the University of British Columbia (UBC).  

%-------------------------------------------------------------------------------
\bibliographystyle{ACM-Reference-Format}
\bibliography{\jobname}

%%% -*-BibTeX-*-
%%% Do NOT edit. File created by BibTeX with style
%%% ACM-Reference-Format-Journals [18-Jan-2012].

\begin{thebibliography}{50}

%%% ====================================================================
%%% NOTE TO THE USER: you can override these defaults by providing
%%% customized versions of any of these macros before the \bibliography
%%% command.  Each of them MUST provide its own final punctuation,
%%% except for \shownote{}, \showDOI{}, and \showURL{}.  The latter two
%%% do not use final punctuation, in order to avoid confusing it with
%%% the Web address.
%%%
%%% To suppress output of a particular field, define its macro to expand
%%% to an empty string, or better, \unskip, like this:
%%%
%%% \newcommand{\showDOI}[1]{\unskip}   % LaTeX syntax
%%%
%%% \def \showDOI #1{\unskip}           % plain TeX syntax
%%%
%%% ====================================================================

\ifx \showCODEN    \undefined \def \showCODEN     #1{\unskip}     \fi
\ifx \showDOI      \undefined \def \showDOI       #1{#1}\fi
\ifx \showISBNx    \undefined \def \showISBNx     #1{\unskip}     \fi
\ifx \showISBNxiii \undefined \def \showISBNxiii  #1{\unskip}     \fi
\ifx \showISSN     \undefined \def \showISSN      #1{\unskip}     \fi
\ifx \showLCCN     \undefined \def \showLCCN      #1{\unskip}     \fi
\ifx \shownote     \undefined \def \shownote      #1{#1}          \fi
\ifx \showarticletitle \undefined \def \showarticletitle #1{#1}   \fi
\ifx \showURL      \undefined \def \showURL       {\relax}        \fi
% The following commands are used for tagged output and should be
% invisible to TeX
\providecommand\bibfield[2]{#2}
\providecommand\bibinfo[2]{#2}
\providecommand\natexlab[1]{#1}
\providecommand\showeprint[2][]{arXiv:#2}

\bibitem[\protect\citeauthoryear{??}{tra}{[n.d.]}]%
        {train-celeb}
 \bibinfo{year}{[n.d.]}\natexlab{}.
\newblock \showarticletitle{CelebA dataset}.
\newblock
\newblock
\shownote{\url{https://github.com/ndb796/CelebA-HQ-Face-Identity-and-Attributes-Recognition-PyTorch
  }.}


\bibitem[\protect\citeauthoryear{??}{acs}{[n.d.]}]%
        {acsac-github}
 \bibinfo{year}{[n.d.]}\natexlab{}.
\newblock \showarticletitle{Code for Februus defense}.
\newblock
\newblock
\shownote{\url{https://github.com/AdelaideAuto-IDLab/Februus.git }.}


\bibitem[\protect\citeauthoryear{??}{str}{[n.d.]}]%
        {strip-github}
 \bibinfo{year}{[n.d.]}\natexlab{}.
\newblock \showarticletitle{Code for STRIP defense}.
\newblock
\newblock
\shownote{\url{https://github.com/garrisongys/STRIP}.}


\bibitem[\protect\citeauthoryear{??}{avg}{[n.d.]}]%
        {avg-filtering}
 \bibinfo{year}{[n.d.]}\natexlab{}.
\newblock \showarticletitle{Image Filtering Median Filtering}.
\newblock
\newblock
\shownote{\url{https://homepages.inf.ed.ac.uk/rbf/HIPR2/mean.htm}.}


\bibitem[\protect\citeauthoryear{??}{ima}{[n.d.]}]%
        {imagenette}
 \bibinfo{year}{[n.d.]}\natexlab{}.
\newblock \showarticletitle{ImageNette dataset}.
\newblock
\newblock
\shownote{\url{https://github.com/fastai/imagenette}.}


\bibitem[\protect\citeauthoryear{Athalye, Carlini, and Wagner}{Athalye
  et~al\mbox{.}}{2018a}]%
        {athalye2018obfuscated}
\bibfield{author}{\bibinfo{person}{Anish Athalye}, \bibinfo{person}{Nicholas
  Carlini}, {and} \bibinfo{person}{David Wagner}.}
  \bibinfo{year}{2018}\natexlab{a}.
\newblock \showarticletitle{Obfuscated gradients give a false sense of
  security: Circumventing defenses to adversarial examples}.
\newblock \bibinfo{journal}{\emph{arXiv preprint arXiv:1802.00420}}
  (\bibinfo{year}{2018}).
\newblock


\bibitem[\protect\citeauthoryear{Athalye, Engstrom, Ilyas, and Kwok}{Athalye
  et~al\mbox{.}}{2018b}]%
        {athalye2018synthesizing}
\bibfield{author}{\bibinfo{person}{Anish Athalye}, \bibinfo{person}{Logan
  Engstrom}, \bibinfo{person}{Andrew Ilyas}, {and} \bibinfo{person}{Kevin
  Kwok}.} \bibinfo{year}{2018}\natexlab{b}.
\newblock \showarticletitle{Synthesizing robust adversarial examples}. In
  \bibinfo{booktitle}{\emph{International conference on machine learning}}.
  PMLR, \bibinfo{pages}{284--293}.
\newblock


\bibitem[\protect\citeauthoryear{Bojarski, Del~Testa, Dworakowski, Firner,
  Flepp, Goyal, Jackel, Monfort, Muller, Zhang, et~al\mbox{.}}{Bojarski
  et~al\mbox{.}}{2016}]%
        {bojarski2016end}
\bibfield{author}{\bibinfo{person}{Mariusz Bojarski}, \bibinfo{person}{Davide
  Del~Testa}, \bibinfo{person}{Daniel Dworakowski}, \bibinfo{person}{Bernhard
  Firner}, \bibinfo{person}{Beat Flepp}, \bibinfo{person}{Prasoon Goyal},
  \bibinfo{person}{Lawrence~D Jackel}, \bibinfo{person}{Mathew Monfort},
  \bibinfo{person}{Urs Muller}, \bibinfo{person}{Jiakai Zhang},
  {et~al\mbox{.}}} \bibinfo{year}{2016}\natexlab{}.
\newblock \showarticletitle{End to end learning for self-driving cars}.
\newblock \bibinfo{journal}{\emph{arXiv preprint arXiv:1604.07316}}
  (\bibinfo{year}{2016}).
\newblock


\bibitem[\protect\citeauthoryear{Brown, Man{\'e}, Roy, Abadi, and Gilmer}{Brown
  et~al\mbox{.}}{2017}]%
        {brown2017adversarial}
\bibfield{author}{\bibinfo{person}{Tom~B Brown}, \bibinfo{person}{Dandelion
  Man{\'e}}, \bibinfo{person}{Aurko Roy}, \bibinfo{person}{Mart{\'\i}n Abadi},
  {and} \bibinfo{person}{Justin Gilmer}.} \bibinfo{year}{2017}\natexlab{}.
\newblock \showarticletitle{Adversarial patch}.
\newblock \bibinfo{journal}{\emph{arXiv preprint arXiv:1712.09665}}
  (\bibinfo{year}{2017}).
\newblock


\bibitem[\protect\citeauthoryear{Cao, Shen, Xie, Parkhi, and Zisserman}{Cao
  et~al\mbox{.}}{2018}]%
        {cao2018vggface2}
\bibfield{author}{\bibinfo{person}{Qiong Cao}, \bibinfo{person}{Li Shen},
  \bibinfo{person}{Weidi Xie}, \bibinfo{person}{Omkar~M Parkhi}, {and}
  \bibinfo{person}{Andrew Zisserman}.} \bibinfo{year}{2018}\natexlab{}.
\newblock \showarticletitle{Vggface2: A dataset for recognising faces across
  pose and age}. In \bibinfo{booktitle}{\emph{2018 13th IEEE international
  conference on automatic face \& gesture recognition (FG 2018)}}. IEEE,
  \bibinfo{pages}{67--74}.
\newblock


\bibitem[\protect\citeauthoryear{Chiang, Ni, Abdelkader, Zhu, Studor, and
  Goldstein}{Chiang et~al\mbox{.}}{2020}]%
        {chiang2020certified}
\bibfield{author}{\bibinfo{person}{Ping-yeh Chiang}, \bibinfo{person}{Renkun
  Ni}, \bibinfo{person}{Ahmed Abdelkader}, \bibinfo{person}{Chen Zhu},
  \bibinfo{person}{Christoph Studor}, {and} \bibinfo{person}{Tom Goldstein}.}
  \bibinfo{year}{2020}\natexlab{}.
\newblock \showarticletitle{Certified defenses for adversarial patches}.
\newblock \bibinfo{journal}{\emph{arXiv preprint arXiv:2003.06693}}
  (\bibinfo{year}{2020}).
\newblock


\bibitem[\protect\citeauthoryear{Chou, Tramer, and Pellegrino}{Chou
  et~al\mbox{.}}{2020}]%
        {chou2020sentinet}
\bibfield{author}{\bibinfo{person}{Edward Chou}, \bibinfo{person}{Florian
  Tramer}, {and} \bibinfo{person}{Giancarlo Pellegrino}.}
  \bibinfo{year}{2020}\natexlab{}.
\newblock \showarticletitle{Sentinet: Detecting localized universal attacks
  against deep learning systems}. In \bibinfo{booktitle}{\emph{2020 IEEE
  Security and Privacy Workshops (SPW)}}. IEEE, \bibinfo{pages}{48--54}.
\newblock


\bibitem[\protect\citeauthoryear{Deng, Dong, Socher, Li, Li, and Fei-Fei}{Deng
  et~al\mbox{.}}{2009}]%
        {deng2009imagenet}
\bibfield{author}{\bibinfo{person}{Jia Deng}, \bibinfo{person}{Wei Dong},
  \bibinfo{person}{Richard Socher}, \bibinfo{person}{Li-Jia Li},
  \bibinfo{person}{Kai Li}, {and} \bibinfo{person}{Li Fei-Fei}.}
  \bibinfo{year}{2009}\natexlab{}.
\newblock \showarticletitle{Imagenet: A large-scale hierarchical image
  database}. In \bibinfo{booktitle}{\emph{2009 IEEE conference on computer
  vision and pattern recognition}}. Ieee, \bibinfo{pages}{248--255}.
\newblock


\bibitem[\protect\citeauthoryear{Doan, Abbasnejad, and Ranasinghe}{Doan
  et~al\mbox{.}}{2020}]%
        {doan2020februus}
\bibfield{author}{\bibinfo{person}{Bao~Gia Doan}, \bibinfo{person}{Ehsan
  Abbasnejad}, {and} \bibinfo{person}{Damith~C Ranasinghe}.}
  \bibinfo{year}{2020}\natexlab{}.
\newblock \showarticletitle{Februus: Input purification defense against trojan
  attacks on deep neural network systems}. In \bibinfo{booktitle}{\emph{Annual
  Computer Security Applications Conference}}. \bibinfo{pages}{897--912}.
\newblock


\bibitem[\protect\citeauthoryear{Esteva, Robicquet, Ramsundar, Kuleshov,
  DePristo, Chou, Cui, Corrado, Thrun, and Dean}{Esteva et~al\mbox{.}}{2019}]%
        {esteva2019guide}
\bibfield{author}{\bibinfo{person}{Andre Esteva}, \bibinfo{person}{Alexandre
  Robicquet}, \bibinfo{person}{Bharath Ramsundar}, \bibinfo{person}{Volodymyr
  Kuleshov}, \bibinfo{person}{Mark DePristo}, \bibinfo{person}{Katherine Chou},
  \bibinfo{person}{Claire Cui}, \bibinfo{person}{Greg Corrado},
  \bibinfo{person}{Sebastian Thrun}, {and} \bibinfo{person}{Jeff Dean}.}
  \bibinfo{year}{2019}\natexlab{}.
\newblock \showarticletitle{A guide to deep learning in healthcare}.
\newblock \bibinfo{journal}{\emph{Nature medicine}} \bibinfo{volume}{25},
  \bibinfo{number}{1} (\bibinfo{year}{2019}), \bibinfo{pages}{24--29}.
\newblock


\bibitem[\protect\citeauthoryear{Eykholt, Evtimov, Fernandes, Li, Rahmati,
  Xiao, Prakash, Kohno, and Song}{Eykholt et~al\mbox{.}}{2018}]%
        {eykholt2018robust}
\bibfield{author}{\bibinfo{person}{Kevin Eykholt}, \bibinfo{person}{Ivan
  Evtimov}, \bibinfo{person}{Earlence Fernandes}, \bibinfo{person}{Bo Li},
  \bibinfo{person}{Amir Rahmati}, \bibinfo{person}{Chaowei Xiao},
  \bibinfo{person}{Atul Prakash}, \bibinfo{person}{Tadayoshi Kohno}, {and}
  \bibinfo{person}{Dawn Song}.} \bibinfo{year}{2018}\natexlab{}.
\newblock \showarticletitle{Robust physical-world attacks on deep learning
  visual classification}. In \bibinfo{booktitle}{\emph{Proceedings of the IEEE
  Conference on Computer Vision and Pattern Recognition}}.
  \bibinfo{pages}{1625--1634}.
\newblock


\bibitem[\protect\citeauthoryear{Gao, Xu, Wang, Chen, Ranasinghe, and
  Nepal}{Gao et~al\mbox{.}}{2019}]%
        {gao2019strip}
\bibfield{author}{\bibinfo{person}{Yansong Gao}, \bibinfo{person}{Change Xu},
  \bibinfo{person}{Derui Wang}, \bibinfo{person}{Shiping Chen},
  \bibinfo{person}{Damith~C Ranasinghe}, {and} \bibinfo{person}{Surya Nepal}.}
  \bibinfo{year}{2019}\natexlab{}.
\newblock \showarticletitle{Strip: A defence against trojan attacks on deep
  neural networks}. In \bibinfo{booktitle}{\emph{Proceedings of the 35th Annual
  Computer Security Applications Conference}}. \bibinfo{pages}{113--125}.
\newblock


\bibitem[\protect\citeauthoryear{Ghorbani, Abid, and Zou}{Ghorbani
  et~al\mbox{.}}{2019}]%
        {ghorbani2019interpretation}
\bibfield{author}{\bibinfo{person}{Amirata Ghorbani}, \bibinfo{person}{Abubakar
  Abid}, {and} \bibinfo{person}{James Zou}.} \bibinfo{year}{2019}\natexlab{}.
\newblock \showarticletitle{Interpretation of neural networks is fragile}. In
  \bibinfo{booktitle}{\emph{Proceedings of the AAAI Conference on Artificial
  Intelligence}}, Vol.~\bibinfo{volume}{33}. \bibinfo{pages}{3681--3688}.
\newblock


\bibitem[\protect\citeauthoryear{Hayes}{Hayes}{2018}]%
        {hayes2018visible}
\bibfield{author}{\bibinfo{person}{Jamie Hayes}.}
  \bibinfo{year}{2018}\natexlab{}.
\newblock \showarticletitle{On visible adversarial perturbations \& digital
  watermarking}. In \bibinfo{booktitle}{\emph{Proceedings of the IEEE
  Conference on Computer Vision and Pattern Recognition Workshops}}.
  \bibinfo{pages}{1597--1604}.
\newblock


\bibitem[\protect\citeauthoryear{Huang, Gao, Zhou, Xie, Yuille, Zou, and
  Liu}{Huang et~al\mbox{.}}{2020}]%
        {huang2020universal}
\bibfield{author}{\bibinfo{person}{Lifeng Huang}, \bibinfo{person}{Chengying
  Gao}, \bibinfo{person}{Yuyin Zhou}, \bibinfo{person}{Cihang Xie},
  \bibinfo{person}{Alan~L Yuille}, \bibinfo{person}{Changqing Zou}, {and}
  \bibinfo{person}{Ning Liu}.} \bibinfo{year}{2020}\natexlab{}.
\newblock \showarticletitle{Universal Physical Camouflage Attacks on Object
  Detectors}. In \bibinfo{booktitle}{\emph{Proceedings of the IEEE/CVF
  Conference on Computer Vision and Pattern Recognition}}.
  \bibinfo{pages}{720--729}.
\newblock


\bibitem[\protect\citeauthoryear{Jha, Raj, Fernandes, Jha, Jha, Verma, Jalaian,
  and Swami}{Jha et~al\mbox{.}}{2019}]%
        {jha2019attribution}
\bibfield{author}{\bibinfo{person}{Susmit Jha}, \bibinfo{person}{Sunny Raj},
  \bibinfo{person}{Steven~Lawrence Fernandes}, \bibinfo{person}{Sumit~Kumar
  Jha}, \bibinfo{person}{Somesh Jha}, \bibinfo{person}{Gunjan Verma},
  \bibinfo{person}{Brian Jalaian}, {and} \bibinfo{person}{Ananthram Swami}.}
  \bibinfo{year}{2019}\natexlab{}.
\newblock \showarticletitle{Attribution-driven causal analysis for detection of
  adversarial examples}.
\newblock \bibinfo{journal}{\emph{arXiv preprint arXiv:1903.05821}}
  (\bibinfo{year}{2019}).
\newblock


\bibitem[\protect\citeauthoryear{Kannan, Kurakin, and Goodfellow}{Kannan
  et~al\mbox{.}}{2018}]%
        {kannan2018adversarial}
\bibfield{author}{\bibinfo{person}{Harini Kannan}, \bibinfo{person}{Alexey
  Kurakin}, {and} \bibinfo{person}{Ian Goodfellow}.}
  \bibinfo{year}{2018}\natexlab{}.
\newblock \showarticletitle{Adversarial logit pairing}.
\newblock \bibinfo{journal}{\emph{arXiv preprint arXiv:1803.06373}}
  (\bibinfo{year}{2018}).
\newblock


\bibitem[\protect\citeauthoryear{Levine and Feizi}{Levine and Feizi}{2020}]%
        {levine2020randomized}
\bibfield{author}{\bibinfo{person}{Alexander Levine} {and}
  \bibinfo{person}{Soheil Feizi}.} \bibinfo{year}{2020}\natexlab{}.
\newblock \showarticletitle{(De) Randomized Smoothing for Certifiable Defense
  against Patch Attacks}.
\newblock \bibinfo{journal}{\emph{arXiv preprint arXiv:2002.10733}}
  (\bibinfo{year}{2020}).
\newblock


\bibitem[\protect\citeauthoryear{Li, Lin, Zhou, Qi, Wang, and Jia}{Li
  et~al\mbox{.}}{2022}]%
        {li2022mat}
\bibfield{author}{\bibinfo{person}{Wenbo Li}, \bibinfo{person}{Zhe Lin},
  \bibinfo{person}{Kun Zhou}, \bibinfo{person}{Lu Qi}, \bibinfo{person}{Yi
  Wang}, {and} \bibinfo{person}{Jiaya Jia}.} \bibinfo{year}{2022}\natexlab{}.
\newblock \showarticletitle{MAT: Mask-Aware Transformer for Large Hole Image
  Inpainting}. In \bibinfo{booktitle}{\emph{Proceedings of the IEEE/CVF
  Conference on Computer Vision and Pattern Recognition}}.
  \bibinfo{pages}{10758--10768}.
\newblock


\bibitem[\protect\citeauthoryear{Li, Wu, Liu, Chen, and Yuan}{Li
  et~al\mbox{.}}{2020}]%
        {li2020advpulse}
\bibfield{author}{\bibinfo{person}{Zhuohang Li}, \bibinfo{person}{Yi Wu},
  \bibinfo{person}{Jian Liu}, \bibinfo{person}{Yingying Chen}, {and}
  \bibinfo{person}{Bo Yuan}.} \bibinfo{year}{2020}\natexlab{}.
\newblock \showarticletitle{AdvPulse: Universal, Synchronization-free, and
  Targeted Audio Adversarial Attacks via Subsecond Perturbations}. In
  \bibinfo{booktitle}{\emph{Proceedings of the 2020 ACM SIGSAC Conference on
  Computer and Communications Security}}. \bibinfo{pages}{1121--1134}.
\newblock


\bibitem[\protect\citeauthoryear{Liu, Reda, Shih, Wang, Tao, and Catanzaro}{Liu
  et~al\mbox{.}}{2018}]%
        {liu2018image}
\bibfield{author}{\bibinfo{person}{Guilin Liu}, \bibinfo{person}{Fitsum~A
  Reda}, \bibinfo{person}{Kevin~J Shih}, \bibinfo{person}{Ting-Chun Wang},
  \bibinfo{person}{Andrew Tao}, {and} \bibinfo{person}{Bryan Catanzaro}.}
  \bibinfo{year}{2018}\natexlab{}.
\newblock \showarticletitle{Image inpainting for irregular holes using partial
  convolutions}. In \bibinfo{booktitle}{\emph{Proceedings of the European
  Conference on Computer Vision (ECCV)}}. \bibinfo{pages}{85--100}.
\newblock


\bibitem[\protect\citeauthoryear{Liu, Luo, Wang, and Tang}{Liu
  et~al\mbox{.}}{2015}]%
        {liu2015faceattributes}
\bibfield{author}{\bibinfo{person}{Ziwei Liu}, \bibinfo{person}{Ping Luo},
  \bibinfo{person}{Xiaogang Wang}, {and} \bibinfo{person}{Xiaoou Tang}.}
  \bibinfo{year}{2015}\natexlab{}.
\newblock \showarticletitle{Deep Learning Face Attributes in the Wild}. In
  \bibinfo{booktitle}{\emph{Proceedings of International Conference on Computer
  Vision (ICCV)}}.
\newblock


\bibitem[\protect\citeauthoryear{McCoyd, Park, Chen, Shah, Roggenkemper, Hwang,
  Liu, and Wagner}{McCoyd et~al\mbox{.}}{2020}]%
        {mccoyd2020minority}
\bibfield{author}{\bibinfo{person}{Michael McCoyd}, \bibinfo{person}{Won Park},
  \bibinfo{person}{Steven Chen}, \bibinfo{person}{Neil Shah},
  \bibinfo{person}{Ryan Roggenkemper}, \bibinfo{person}{Minjune Hwang},
  \bibinfo{person}{Jason~Xinyu Liu}, {and} \bibinfo{person}{David Wagner}.}
  \bibinfo{year}{2020}\natexlab{}.
\newblock \showarticletitle{Minority Reports Defense: Defending Against
  Adversarial Patches}.
\newblock \bibinfo{journal}{\emph{arXiv preprint arXiv:2004.13799}}
  (\bibinfo{year}{2020}).
\newblock


\bibitem[\protect\citeauthoryear{Mundhenk, Chen, and Friedland}{Mundhenk
  et~al\mbox{.}}{2019}]%
        {mundhenk2019efficient}
\bibfield{author}{\bibinfo{person}{T~Nathan Mundhenk}, \bibinfo{person}{Barry~Y
  Chen}, {and} \bibinfo{person}{Gerald Friedland}.}
  \bibinfo{year}{2019}\natexlab{}.
\newblock \showarticletitle{Efficient saliency maps for explainable AI}.
\newblock \bibinfo{journal}{\emph{arXiv preprint arXiv:1911.11293}}
  (\bibinfo{year}{2019}).
\newblock


\bibitem[\protect\citeauthoryear{Naseer, Khan, and Porikli}{Naseer
  et~al\mbox{.}}{2019}]%
        {naseer2019local}
\bibfield{author}{\bibinfo{person}{Muzammal Naseer}, \bibinfo{person}{Salman
  Khan}, {and} \bibinfo{person}{Fatih Porikli}.}
  \bibinfo{year}{2019}\natexlab{}.
\newblock \showarticletitle{Local gradients smoothing: Defense against
  localized adversarial attacks}. In \bibinfo{booktitle}{\emph{2019 IEEE Winter
  Conference on Applications of Computer Vision (WACV)}}. IEEE,
  \bibinfo{pages}{1300--1307}.
\newblock


\bibitem[\protect\citeauthoryear{Parkhi, Vedaldi, and Zisserman}{Parkhi
  et~al\mbox{.}}{2015}]%
        {parkhi2015deep}
\bibfield{author}{\bibinfo{person}{Omkar~M Parkhi}, \bibinfo{person}{Andrea
  Vedaldi}, {and} \bibinfo{person}{Andrew Zisserman}.}
  \bibinfo{year}{2015}\natexlab{}.
\newblock \showarticletitle{Deep face recognition}.
\newblock  (\bibinfo{year}{2015}).
\newblock


\bibitem[\protect\citeauthoryear{Pathak, Krahenbuhl, Donahue, Darrell, and
  Efros}{Pathak et~al\mbox{.}}{2016}]%
        {pathak2016context}
\bibfield{author}{\bibinfo{person}{Deepak Pathak}, \bibinfo{person}{Philipp
  Krahenbuhl}, \bibinfo{person}{Jeff Donahue}, \bibinfo{person}{Trevor
  Darrell}, {and} \bibinfo{person}{Alexei~A Efros}.}
  \bibinfo{year}{2016}\natexlab{}.
\newblock \showarticletitle{Context encoders: Feature learning by inpainting}.
  In \bibinfo{booktitle}{\emph{Proceedings of the IEEE conference on computer
  vision and pattern recognition}}. \bibinfo{pages}{2536--2544}.
\newblock


\bibitem[\protect\citeauthoryear{Rao and Frtunikj}{Rao and Frtunikj}{2018}]%
        {rao2018deep}
\bibfield{author}{\bibinfo{person}{Qing Rao} {and} \bibinfo{person}{Jelena
  Frtunikj}.} \bibinfo{year}{2018}\natexlab{}.
\newblock \showarticletitle{Deep learning for self-driving cars: Chances and
  challenges}. In \bibinfo{booktitle}{\emph{Proceedings of the 1st
  International Workshop on Software Engineering for AI in Autonomous
  Systems}}. \bibinfo{pages}{35--38}.
\newblock


\bibitem[\protect\citeauthoryear{Rao, Stutz, and Schiele}{Rao
  et~al\mbox{.}}{2020}]%
        {rao2020adversarial}
\bibfield{author}{\bibinfo{person}{Sukrut Rao}, \bibinfo{person}{David Stutz},
  {and} \bibinfo{person}{Bernt Schiele}.} \bibinfo{year}{2020}\natexlab{}.
\newblock \showarticletitle{Adversarial Training against Location-Optimized
  Adversarial Patches}.
\newblock \bibinfo{journal}{\emph{arXiv preprint arXiv:2005.02313}}
  (\bibinfo{year}{2020}).
\newblock


\bibitem[\protect\citeauthoryear{Sahiner, Pezeshk, Hadjiiski, Wang, Drukker,
  Cha, Summers, and Giger}{Sahiner et~al\mbox{.}}{2019}]%
        {sahiner2019deep}
\bibfield{author}{\bibinfo{person}{Berkman Sahiner}, \bibinfo{person}{Aria
  Pezeshk}, \bibinfo{person}{Lubomir~M Hadjiiski}, \bibinfo{person}{Xiaosong
  Wang}, \bibinfo{person}{Karen Drukker}, \bibinfo{person}{Kenny~H Cha},
  \bibinfo{person}{Ronald~M Summers}, {and} \bibinfo{person}{Maryellen~L
  Giger}.} \bibinfo{year}{2019}\natexlab{}.
\newblock \showarticletitle{Deep learning in medical imaging and radiation
  therapy}.
\newblock \bibinfo{journal}{\emph{Medical physics}} \bibinfo{volume}{46},
  \bibinfo{number}{1} (\bibinfo{year}{2019}), \bibinfo{pages}{e1--e36}.
\newblock


\bibitem[\protect\citeauthoryear{Selvaraju, Cogswell, Das, Vedantam, Parikh,
  and Batra}{Selvaraju et~al\mbox{.}}{2017}]%
        {selvaraju2017grad}
\bibfield{author}{\bibinfo{person}{Ramprasaath~R Selvaraju},
  \bibinfo{person}{Michael Cogswell}, \bibinfo{person}{Abhishek Das},
  \bibinfo{person}{Ramakrishna Vedantam}, \bibinfo{person}{Devi Parikh}, {and}
  \bibinfo{person}{Dhruv Batra}.} \bibinfo{year}{2017}\natexlab{}.
\newblock \showarticletitle{Grad-cam: Visual explanations from deep networks
  via gradient-based localization}. In \bibinfo{booktitle}{\emph{Proceedings of
  the IEEE international conference on computer vision}}.
  \bibinfo{pages}{618--626}.
\newblock


\bibitem[\protect\citeauthoryear{Smilkov, Thorat, Kim, Vi{\'e}gas, and
  Wattenberg}{Smilkov et~al\mbox{.}}{2017}]%
        {smilkov2017smoothgrad}
\bibfield{author}{\bibinfo{person}{Daniel Smilkov}, \bibinfo{person}{Nikhil
  Thorat}, \bibinfo{person}{Been Kim}, \bibinfo{person}{Fernanda Vi{\'e}gas},
  {and} \bibinfo{person}{Martin Wattenberg}.} \bibinfo{year}{2017}\natexlab{}.
\newblock \showarticletitle{Smoothgrad: removing noise by adding noise}.
\newblock \bibinfo{journal}{\emph{arXiv preprint arXiv:1706.03825}}
  (\bibinfo{year}{2017}).
\newblock


\bibitem[\protect\citeauthoryear{Song, Yu, Peng, and Narasimhan}{Song
  et~al\mbox{.}}{2020}]%
        {song2020universal}
\bibfield{author}{\bibinfo{person}{Liwei Song}, \bibinfo{person}{Xinwei Yu},
  \bibinfo{person}{Hsuan-Tung Peng}, {and} \bibinfo{person}{Karthik
  Narasimhan}.} \bibinfo{year}{2020}\natexlab{}.
\newblock \showarticletitle{Universal Adversarial Attacks with Natural Triggers
  for Text Classification}.
\newblock \bibinfo{journal}{\emph{arXiv preprint arXiv:2005.00174}}
  (\bibinfo{year}{2020}).
\newblock


\bibitem[\protect\citeauthoryear{Sun, Liang, Wang, and Tang}{Sun
  et~al\mbox{.}}{2015}]%
        {sun2015deepid3}
\bibfield{author}{\bibinfo{person}{Yi Sun}, \bibinfo{person}{Ding Liang},
  \bibinfo{person}{Xiaogang Wang}, {and} \bibinfo{person}{Xiaoou Tang}.}
  \bibinfo{year}{2015}\natexlab{}.
\newblock \showarticletitle{Deepid3: Face recognition with very deep neural
  networks}.
\newblock \bibinfo{journal}{\emph{arXiv preprint arXiv:1502.00873}}
  (\bibinfo{year}{2015}).
\newblock


\bibitem[\protect\citeauthoryear{Sundararajan, Taly, and Yan}{Sundararajan
  et~al\mbox{.}}{2017}]%
        {sundararajan2017axiomatic}
\bibfield{author}{\bibinfo{person}{Mukund Sundararajan}, \bibinfo{person}{Ankur
  Taly}, {and} \bibinfo{person}{Qiqi Yan}.} \bibinfo{year}{2017}\natexlab{}.
\newblock \showarticletitle{Axiomatic attribution for deep networks}.
\newblock \bibinfo{journal}{\emph{arXiv preprint arXiv:1703.01365}}
  (\bibinfo{year}{2017}).
\newblock


\bibitem[\protect\citeauthoryear{Szegedy, Zaremba, Sutskever, Bruna, Erhan,
  Goodfellow, and Fergus}{Szegedy et~al\mbox{.}}{2013}]%
        {szegedy2013intriguing}
\bibfield{author}{\bibinfo{person}{Christian Szegedy},
  \bibinfo{person}{Wojciech Zaremba}, \bibinfo{person}{Ilya Sutskever},
  \bibinfo{person}{Joan Bruna}, \bibinfo{person}{Dumitru Erhan},
  \bibinfo{person}{Ian Goodfellow}, {and} \bibinfo{person}{Rob Fergus}.}
  \bibinfo{year}{2013}\natexlab{}.
\newblock \showarticletitle{Intriguing properties of neural networks}.
\newblock \bibinfo{journal}{\emph{arXiv preprint arXiv:1312.6199}}
  (\bibinfo{year}{2013}).
\newblock


\bibitem[\protect\citeauthoryear{Wu, Tong, and Vorobeychik}{Wu
  et~al\mbox{.}}{2020}]%
        {wu2020defending}
\bibfield{author}{\bibinfo{person}{Tong Wu}, \bibinfo{person}{Liang Tong},
  {and} \bibinfo{person}{Yevgeniy Vorobeychik}.}
  \bibinfo{year}{2020}\natexlab{}.
\newblock \showarticletitle{Defending Against Physically Realizable Attacks on
  Image Classification}. In \bibinfo{booktitle}{\emph{International Conference
  on Learning Representations}}.
\newblock


\bibitem[\protect\citeauthoryear{Xiang, Bhagoji, Sehwag, and Mittal}{Xiang
  et~al\mbox{.}}{2020}]%
        {xiang2020patchguard}
\bibfield{author}{\bibinfo{person}{Chong Xiang}, \bibinfo{person}{Arjun~Nitin
  Bhagoji}, \bibinfo{person}{Vikash Sehwag}, {and} \bibinfo{person}{Prateek
  Mittal}.} \bibinfo{year}{2020}\natexlab{}.
\newblock \showarticletitle{PatchGuard: Provable Defense against Adversarial
  Patches Using Masks on Small Receptive Fields}.
\newblock \bibinfo{journal}{\emph{arXiv preprint arXiv:2005.10884}}
  (\bibinfo{year}{2020}).
\newblock


\bibitem[\protect\citeauthoryear{Xiang, Mahloujifar, and Mittal}{Xiang
  et~al\mbox{.}}{2021}]%
        {xiang2021patchcleanser}
\bibfield{author}{\bibinfo{person}{Chong Xiang}, \bibinfo{person}{Saeed
  Mahloujifar}, {and} \bibinfo{person}{Prateek Mittal}.}
  \bibinfo{year}{2021}\natexlab{}.
\newblock \showarticletitle{PatchCleanser: Certifiably Robust Defense against
  Adversarial Patches for Any Image Classifier}.
\newblock \bibinfo{journal}{\emph{arXiv preprint arXiv:2108.09135}}
  (\bibinfo{year}{2021}).
\newblock


\bibitem[\protect\citeauthoryear{Xie, Tan, Gong, Yuille, and Le}{Xie
  et~al\mbox{.}}{2020}]%
        {xie2020smooth}
\bibfield{author}{\bibinfo{person}{Cihang Xie}, \bibinfo{person}{Mingxing Tan},
  \bibinfo{person}{Boqing Gong}, \bibinfo{person}{Alan Yuille}, {and}
  \bibinfo{person}{Quoc~V Le}.} \bibinfo{year}{2020}\natexlab{}.
\newblock \showarticletitle{Smooth adversarial training}.
\newblock \bibinfo{journal}{\emph{arXiv preprint arXiv:2006.14536}}
  (\bibinfo{year}{2020}).
\newblock


\bibitem[\protect\citeauthoryear{Xu, Yu, and Chen}{Xu et~al\mbox{.}}{2020}]%
        {xu2020lance}
\bibfield{author}{\bibinfo{person}{Zirui Xu}, \bibinfo{person}{Fuxun Yu}, {and}
  \bibinfo{person}{Xiang Chen}.} \bibinfo{year}{2020}\natexlab{}.
\newblock \showarticletitle{LanCe: A comprehensive and lightweight CNN defense
  methodology against physical adversarial attacks on embedded multimedia
  applications}. In \bibinfo{booktitle}{\emph{2020 25th Asia and South Pacific
  Design Automation Conference (ASP-DAC)}}. IEEE, \bibinfo{pages}{470--475}.
\newblock


\bibitem[\protect\citeauthoryear{Yu, Lin, Yang, Shen, Lu, and Huang}{Yu
  et~al\mbox{.}}{2019}]%
        {yu2019free}
\bibfield{author}{\bibinfo{person}{Jiahui Yu}, \bibinfo{person}{Zhe Lin},
  \bibinfo{person}{Jimei Yang}, \bibinfo{person}{Xiaohui Shen},
  \bibinfo{person}{Xin Lu}, {and} \bibinfo{person}{Thomas~S Huang}.}
  \bibinfo{year}{2019}\natexlab{}.
\newblock \showarticletitle{Free-form image inpainting with gated convolution}.
  In \bibinfo{booktitle}{\emph{Proceedings of the IEEE International Conference
  on Computer Vision}}. \bibinfo{pages}{4471--4480}.
\newblock


\bibitem[\protect\citeauthoryear{Zhang, Wang, Shen, Ji, Luo, and Wang}{Zhang
  et~al\mbox{.}}{2020}]%
        {zhang2020interpretable}
\bibfield{author}{\bibinfo{person}{Xinyang Zhang}, \bibinfo{person}{Ningfei
  Wang}, \bibinfo{person}{Hua Shen}, \bibinfo{person}{Shouling Ji},
  \bibinfo{person}{Xiapu Luo}, {and} \bibinfo{person}{Ting Wang}.}
  \bibinfo{year}{2020}\natexlab{}.
\newblock \showarticletitle{Interpretable deep learning under fire}. In
  \bibinfo{booktitle}{\emph{29th $\{$USENIX$\}$ Security Symposium
  ($\{$USENIX$\}$ Security 20)}}.
\newblock


\bibitem[\protect\citeauthoryear{Zheng, Cham, and Cai}{Zheng
  et~al\mbox{.}}{2019}]%
        {zheng2019pluralistic}
\bibfield{author}{\bibinfo{person}{Chuanxia Zheng}, \bibinfo{person}{Tat-Jen
  Cham}, {and} \bibinfo{person}{Jianfei Cai}.} \bibinfo{year}{2019}\natexlab{}.
\newblock \showarticletitle{Pluralistic image completion}. In
  \bibinfo{booktitle}{\emph{Proceedings of the IEEE Conference on Computer
  Vision and Pattern Recognition}}. \bibinfo{pages}{1438--1447}.
\newblock


\bibitem[\protect\citeauthoryear{Zhou, Lapedriza, Khosla, Oliva, and
  Torralba}{Zhou et~al\mbox{.}}{2017}]%
        {zhou2017places}
\bibfield{author}{\bibinfo{person}{Bolei Zhou}, \bibinfo{person}{Agata
  Lapedriza}, \bibinfo{person}{Aditya Khosla}, \bibinfo{person}{Aude Oliva},
  {and} \bibinfo{person}{Antonio Torralba}.} \bibinfo{year}{2017}\natexlab{}.
\newblock \showarticletitle{Places: A 10 million Image Database for Scene
  Recognition}.
\newblock \bibinfo{journal}{\emph{IEEE Transactions on Pattern Analysis and
  Machine Intelligence}} (\bibinfo{year}{2017}).
\newblock


\end{thebibliography}

\pagebreak
\appendix  
\section{Appendix} 
\label{sec:appendix}

 \begin{figure}[t]
\centering
  \includegraphics[ height=1.0in,trim=3 3 3 3,clip]{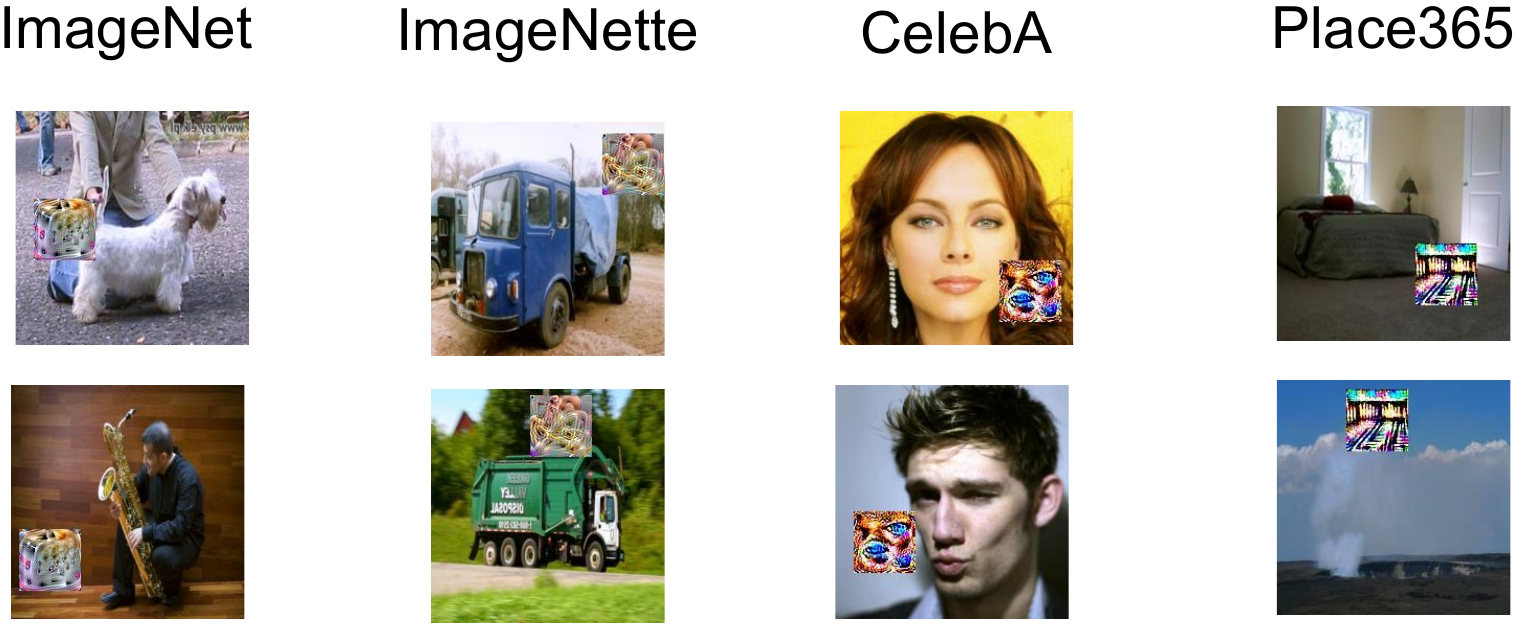}
\caption{Adversarial examples on each dataset. }
\label{fig:sample-img}
%\vspace{-4mm}
\end{figure}

\subsection{{\sysname's Performance under Different Parameter Values}}
\label{sec:select-parameters}
{
As mentioned in Section~\ref{sec:technique-setup}, \sysname has three main parameters in its defense setup: (1) kernel size for pre-processing (average filtering) the saliency map; (2) size of the detection bounding box; and (3) number of hold-out images used for feature transplantation. 
This section evaluates \sysname's performance by varying each of these parameters (all on ImageNet), from which we choose the parameters used in our main evaluation.

\subsubsection{Kernel size for pre-processing (average filtering) the saliency map}
We consider the following different kernel sizes for pre-processing the saliency map: 11, 31, 51 and 71, and the results are in Table~\ref{tab:diff-kernel-sizes}. 

When the kernel size is small, \sysname has lower detection recall (93.28\% for kernel size of 11 vs. over 97\% for other larger sizes). 
Low detection recall means \sysname erroneously identifies many natural (benign) features as the suspicious features in adversarial examples (hence the actual adversarial patch remains undetected). 
This is because the natural features would still remain as the salient feature if the kernel size is small (i.e., the kernel area for average filtering).
Instead, a larger kernel size can smooth out the regions associated with the benign features, and the regions associated with adversarial patch still remain salient and can be identified by \sysname.  
Based on the above analysis, we use a kernel size of 51 in our main evaluation (which has a good balance between different performance metrics).

\begin{table}[H]
\begin{threeparttable}
\caption{\sysname's performance by using different kernel sizes for pre-processing the saliency map. }

\label{tab:diff-kernel-sizes}
\centering  
\footnotesize

\begin{tabular}{|c| cccc |}
\hline
\multirow{2}{12em}{\centering \textbf{Metric} (\%)}  & \multicolumn{4}{|c|}{\textbf{Kernel size for average filtering}}  \\

%\cline{3-10}
  &  11 & 31 & 51 & 71   \\
\hline

\textbf{Mitigation Success Recall} & 93.28 & 97.28 & 98.35 & 98.04  \\
\hline
\textbf{Robust Accuracy}   & 71.37 & 74.53 & 76.14 & 75.78   \\
\hline
\textbf{False Positive}    & 1.38 & 1.6 & 1.87 & 1.87    \\

\hline
\end{tabular}

\end{threeparttable}
\vspace{-4mm}
\end{table}

\subsubsection{Size of the detection bounding box}
We consider the following different sizes of the detection bounding box: 78, 90, 102, 114, 122, which correspond to around 12\%, 16\%, 20\%, 25\% and 30\% image pixels. 
We report the results in Table~\ref{tab:diff-bounding-box}. 

When the detection bounding box is small, it is more difficult for \sysname to locate the adversarial patch from the adversarial examples, which leads to lower detection recall as shown. Conversely, we see that detection recall increases as the detection bounding box grows in size. 

On the other hand, larger bounding box could also degrade robust accuracy and cause more FP. This is because when the bounding box is large, more contents need to be recovered by the GAN, which is a more challenging task.

Based on the above, bounding boxes in different sizes can be used based on different objectives, e.g., one can use a smaller bounding box to minimize FP while maintaining good detection and mitigation performance.
We use a size of 102 in our main evaluation.

}

\begin{table}[H]
\begin{threeparttable}
\caption{\sysname's performance by using detection bounding boxes in different sizes}

\label{tab:diff-bounding-box}
\centering  
\footnotesize

\begin{tabular}{|c| ccccc |}
\hline
\multirow{2}{12em}{\centering \textbf{Metric} (\%)}  & \multicolumn{5}{|c|}{\textbf{Size of detection bounding box}}  \\

%\cline{3-10}
  &  78 & 90 & 102 & 114 & 122  \\
\hline

\textbf{Mitigation Success Recall} & 91.93  & 94.89 & 97.66 & 98.93 & 99.07 \\
\hline

\textbf{Robust Accuracy}   & 81.88 & 80.89 & 78.69 & 70.35 & 63.13  \\
\hline
\textbf{False Positive}    & 0.08 & 0.7 & 1.66 & 4.34 & 7.82   \\

\hline
\end{tabular}

\end{threeparttable}
\vspace{-2mm}
\end{table}

\subsubsection{{Number of hold-out images used for feature transplantation} }
\label{sec:diff-img-detection}
This section shows \sysname's performance when using different number of hold-out images for feature transplantation (in attack detection).
The results are presented in Table~\ref{tab:detection-with-diff-img}.

As shown, detection success recall reduces as the number of images for detection increases. 
This is because \sysname determines an adversarial example only if \emph{all} the images implanted with the salient features have the same labels as the original test image.
When more images are used for detection, it is more difficult for the adversarial patch to cause the same misclassification on all the hold-out images (it is easier to cause the targeted misclassification on 1 image than on 2 images or more).
Hence, \sysname's detection performance degrades when more images are used for detection.

We also observe that the FPR reduces as we use more images for detection. 
The reason is similar to the above.
Specifically, a benign image will be mis-detected only if its salient features cause the same prediction labels on all the hold-out images after feature transplantation, which is increasingly difficult as the number of hold-out images increase.

Based on the above, different number of hold-out images can be used based on different objectives, e.g., one can use more images for detection in order to minimize FP.
We use 2 images for detection in our main evaluation.

\begin{table}[H] 
	\caption{\sysname's performance when using different number of hold-out images for detection.}
	\label{tab:detection-with-diff-img}
	\centering  
	\footnotesize
	\begin{tabular}{|c| ccc |}
		\hline
		\multirow{2}{*}{\textbf{Metric} (\%)}  & \multicolumn{3}{|c|}{\textbf{Number of images for detection}}  \\
		
		&  1 & 2 & 3   \\
		\hline

		{\textbf{Mitigation Success Recall}} &  98.79 & 97.58 & 96.40       \\ 
		
		\hline
		{\textbf{Robust Accuracy}}   &  78.59 & 78.20 & 77.00    \\                                                                
		\hline
		{\textbf{False Positive}}   &  4.78 & 1.76 & 0.82    \\                                                                
		\hline

	\end{tabular}
	\vspace{-2mm}
\end{table}

\subsection{{Ablation Study}}
\label{sec:ablation-study} 
We consider the following four components in the ablation study. 

(1) \emph{Pre-processing the saliency map to identify suspicious features (for attack detection)}.
We compare \sysname's performance with and without the pre-processing component (on ImageNet with 10\% patch). 
Without the pre-processing component, \sysname detects only 91.93\% AEs (Vs. 98.35\% with the pre-processing component), because it cannot correctly locate the adversarial patch region from the AEs (instead it identifies many natural features as the suspicious features). 
The low detection performance also leads to lower robust accuracy of 71.99\% Vs. 76.14\% with the pre-processing component.

(2) \emph{Guided feature transplantation to reduce FPRs.} 
We compare the performance of using guided feature transplantation Vs. random feature transplantation. 
While both approaches achieve similar detection recall (98.35\% and 99.07\%) and robust accuracy (75.11\% and 76.22\%), the proposed guided feature transplantation achieves an FPR of 1.55\% Vs. 2.9\% by the random transplantation, which constitutes a reduction of FPR by 46.6\%.

(3) \emph{GAN-based attack recovery}. 
We compare the GAN-based and masking-alone attack mitigation strategy, and the detailed results are reported and discussed earlier in Section~\ref{sec:mitigation-eval}. We summarize the main difference here. 
The proposed GAN-based recovery  method outperforms the basic masking-alone strategy (for 75\% and 100\% masking) with: (1) higher robust accuracy (6.56\% higher for 100\% masking); and (2) lower FPRs (0.25\% lower),  because masking alone will cause the loss of semantic contents, which is undesirable for prediction comparison, while the GAN can recover the semantic contents from the masked pixels\footnote{Note that the GAN is \emph{not} always able to recover the correct contents, which explains why the robust accuracy is not as high as the detection recall. Nevertheless, the proposed GAN-based recovery \emph{still outperforms} the basic masking-alone strategy, and is hence adopted by \sysname.
}. 

(4) \emph{Prediction comparison for reducing FPRs.} 
Table~\ref{tab:detection} shows that \sysname yields an average FPR of 3.33\%, which can be reduced to 0.5\%$\sim$0.71\% (depending on different masking percentages used - see Table~\ref{tab:mitigation}). 
By comparing the prediction label on the original and recovered inputs using GAN, \sysname can effectively reduce FP on benign examples, which enables a FPR reduction by 78\%$\sim$85\%. 

In summary, the ablation study shows that the above components in \sysname are crucial in enabling \sysname to achieve high detection performance, high robust accuracy and low FPRs.

\subsection{{Extending \sysname to Defend Against Multi-patch Attacks}}
\label{sec:multi-patch}
This section evaluated the extension of \sysname against multi-patch attacks as discussed in Section~\ref{sec:limitation}. 
To do so, we modify \sysname to \emph{iteratively} perform detection and mitigation until only the benign features are left in the images, i.e., the current suspicious features fail to cause misclassification on the hold-out inputs. 
We consider both 2-patch and 3-patch attacks, and an example is shown below. 
We report the results in Table~\ref{tab:multi-patch}.

\begin{wrapfigure}{R}{0.3\textwidth}
\centering
\includegraphics[width=0.25\textwidth]{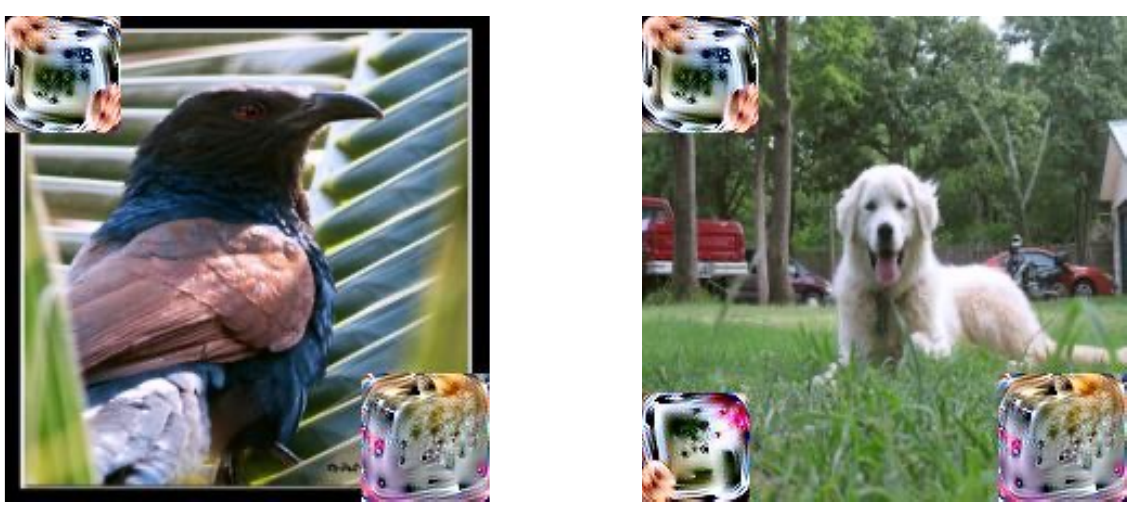}
\label{fig:multi-patch}
\vspace{-2mm}
\end{wrapfigure}

With the above extension, we can see that \sysname still  remains as an effective defense against patch attacks.
In both cases, \sysname detects over 96\% adversarial samples with around 2\% FPR. 
For 2-patch attack, \sysname still achieves a very high robust accuracy of 73.11\%, which is slightly lower than that of 77.47\% on single-patch attack. 
For 3-patch attack, however, \sysname achieves a much lower robust accuracy of 46.98\%, which is because mitigating 3 patches in one single image is more difficult than single- or two-patch attacks. 
Our result therefore demonstrates that \sysname can be extended to effectively defend against multi-patch attack.

\begin{table}[H]
\begin{threeparttable}
\caption{\sysname's performance against multi-patch attacks}

\label{tab:multi-patch}
%\centering  
\footnotesize

\begin{tabular}{|c| c | c |}
\hline
\multirow{1}{5em}{\textbf{Metric} (\%)}  &  \multirow{1}{8em}{\centering \textbf{2-patch attack}} &  \multirow{1}{8em}{\centering \textbf{3-patch attack}} \\

\hline

\textbf{Mitigation Success Recall} & 96.16  & 96.78  \\
\hline

\textbf{Robust Accuracy}   & 73.11 & 46.98   \\
\hline
\textbf{False Positive}    & 2.26 & 1.93   \\

\hline
\end{tabular}

\end{threeparttable}
\vspace{-2mm}
\end{table}

\subsection{{{Extending \sysname to Defend Against Rectangular Patch Attacks} }}
\label{sec:diff-patch-shape} 

{
We now evaluate how \sysname can be extended to defend against rectangular patch attacks.
We train the rectangular patches on each dataset using a 7\% patch (36*96). 
As discussed in Section~\ref{sec:limitation}, we assume the defender is aware of the shape of the patch and hence we use a rectangular bounding box, which occupies around 20\% of pixels as before, and it has a width/height ratio of 6:4. 
Table~\ref{tab:defend-rect-patch} shows the results on different datasets. 
On average, \sysname is able to detect 97.25\% adversarial examples, achieve robust accuracy of 78.3\% with only 0.57\% false positive.

{The current extension assumes the defender's knowledge of the potential patch shape, and we leave the improvement of \sysname to be general to different patch shape in future work. 
For example, instead of using one single detection bounding box, can we use multiple detection boxes in different shapes to cover different potential patches?
}

\begin{table}[H]
\caption{\sysname's performance on \emph{rectangular} patch attack. }
\label{tab:defend-rect-patch}
\centering  
\footnotesize
\begin{tabular}{|c| c | c| c | c  | c |}
\hline

{\textbf{Metric} (\%)}  &  {\textbf{ImageNet}} & {\textbf{ImageNette}} & {\textbf{CelebA}} & {\textbf{Place365}}   & {\textbf{Average}}  \\        

\hline 

 {\textbf{Robust Accuracy}}   & 75.44 & 92.42 & 64.58 & 80.76  &  78.3  \\ 

\hline 
 {\textbf{False Positive}}   &  1.54 & 0.67 & 0.00  & 0.06 & 0.57 \\ 

\hline 

\textbf{Mitigation} & \multirow{2}{1.3em}{98.43} & \multirow{2}{1.3em}{98.84} & \multirow{2}{1.3em}{92.43} & \multirow{2}{1.3em}{99.30} & \multirow{2}{1.3em}{97.25}\\            
\textbf{Success Recall} & & & & & \\

\hline
\end{tabular}
\vspace{-2mm}
\end{table}

\subsection{{Extending \sysname to Defend Against Untargeted Attacks}}
\label{sec:untargeted}
We now describe our effort in modifying \sysname against untargeted attacks. 
\sysname's current attack detection is not designed for untargeted attacks, and hence we modify \sysname by directly performing attack mitigation on the suspicious features (the procedure to identify the suspicious features is the same), and using the use the prediction label on the recovered image as the final output.  

Since we do not use \sysname's detection, we focus on the robust accuracy and FPR on benign sample as the evaluation metric. 
In this setting, \sysname achieves a robust accuracy of 55.15\%, but with an elevated FPR of 44\%. 
\sysname still achieves over 55\% robust accuracy because it is able to successfully identify the adversarial patch region in many AEs (over 68\%) and hence it can perform attack recovery on these adversarial patch regions. 
On the other hand, \sysname yields a high FPR because \sysname does not incorporate its attack detection pipeline to distinguish benign samples and directly uses our mitigation strategy to reduce FPR (Section~\ref{sec:reduce-fp}), which can only be reduced to 44\%. 

Therefore, future work could look into combining \sysname with other defenses, e.g., recent efforts use pre-defined thresholds~\cite{naseer2019local} or performs small masking on the test image and use the prediction disagreement on the masked images~\cite{xiang2021patchcleanser} to characterize benign and adversarial examples, which may be combined with \sysname to facilitate an effective attack detection and mitigation defense.

\subsection{{Comparison with 2 Trojan-attack Defenses}}
\label{sec:trojan-defense-comp}
In Section~\ref{sec:comparison}, we compare with 4 existing defenses designed for countering patch attacks.
We now evaluate 2 additional defenses built for trojan attacks (STRIP~\cite{gao2019strip} and Februus~\cite{doan2020februus}), and we evaluate whether they are also effective against patch attacks. 
We choose these two techniques because: 
(1) STRIP relies on superimposing two different images to detect trojan attack, which may also be effective for patch attacks if the adversarial examples (corrupted with adversarial patch) continue to cause misclassification after being superimposed with another image. 
(2) Februus uses pre-defined threshold to scan the saliency map for trojan attack detection, which is similar to other detection techniques for patch attacks~\cite{chou2020sentinet,naseer2019local}.  

\emph{Comparison with STRIP~\cite{gao2019strip}.}
Gao et al.~\cite{gao2019strip} propose STRIP to defend  against patch-like trojaned adversarial examples by superimposing the entire target image with a number of new images, and detect adversarial examples based on the prediction entropy on the set of new images. 
The prediction entropy is compared against a detection boundary (derived from benign inputs), and a low entropy indicates that the target image is adversarial. 
We use the implementation from \cite{strip-github}, and we use 2000 images for deriving the detection threshold and construct 100 superimposed examples per testing image, similar to the original paper. 

We conduct the evaluation on ImageNet, and found that STRIP only detects around 6\% of the adversarial examples while \sysname detects over 99\% (The FPRs by both techniques are both less than 2\%).  
STRIP achieves low detection performance because the adversarial patch is no longer effective after being blended with the new images, thus the prediction entropy is high on the new images.

\emph{Comparison with {Februus~\cite{doan2020februus}.}}
Doan et al.~\cite{doan2020februus} propose Februus to defend against patch-like trojaned adversarial examples, which first performs attack detection by identifying the regions that exceed pre-defined threshold in the saliency map, and then performs image restoration on these regions for attack mitigation. 
We use the original implementation from \cite{acsac-github} on VGGFace2~\cite{cao2018vggface2}.

Under the VGGFace2 dataset, \sysname achieves a robust accuracy of 37.26\%\footnote{{\sysname has a lower robust accuracy on VGGFace2 than those on the other datasets due to the insufficient performance yielded by the GAN (PICNet~\cite{zheng2019pluralistic}).
This is because we need to train the PICNet from scratch on VGGFace2, which is very time-consuming as VGGFace2 is a very large dataset. 
We trained the PICNet on a small subset of the dataset for a week and used it in our evaluation due to time constraint.
The performance of \sysname can be further improved with more resources to train the PICNet (e.g., increase the size of training set and number of epoches).}}, which is significantly higher than that of 0.2\%\footnote{{To ensure the code was implemented correctly, we verified that the code was able to reproduce the results reported in the original paper for trojan attack. We then used the code to evaluate against patch attacks.}} by Februus. 
This is because Februus relies on the pre-defined threshold to identify the regions associated with adversarial patch. 
This method would fail to locate the adversarial patch if the patch's influence to the prediction is lower than the threshold, and our experiment validates this.

Our results show that although STRIP~\cite{gao2019strip} and Februus~\cite{doan2020februus} are effective defenses against trojan attack, they are {not} able to defend against patch attacks.

\begin{figure}[t]
\centering
  \includegraphics[ height=1.6in ]{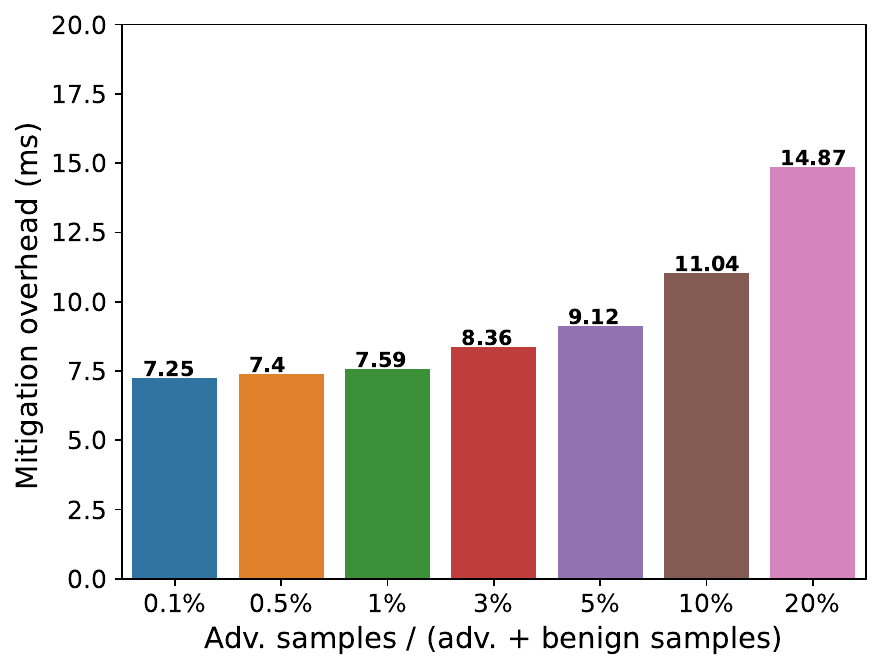}
\caption{Overhead of performing attack mitigation by \sysname, under different ratios of adversarial examples within all the test inputs. One inference pass on the undefended model took 5.56\emph{ms}. 
}
\label{fig:overhead} 
\vspace{-2mm}
\end{figure}

\subsection{Overhead of \sysname}
\label{sec:overhead}

\emph{Attack detection.}
\sysname's detection involves three steps: (1) locate the salient features from the saliency map; (2) identify the least-salient region of the hold-out inputs and perform feature transplantation and (3) prediction comparison. 
We perform the evaluation on the ImageNet dataset, and repeat the evaluation 5 times and report the average overhead (on a single Nvidia RTX 3090 GPU).

In step 2, the identification of the least-salient region of the hold-out inputs can be performed offline (since it is independent of the runtime input), hence we evaluate the overhead in performing feature transplantation only. 
Step 3 requires prediction comparison on 3 images (1 original image and 2 hold-out images implanted with suspicious features), which can be executed in parallel to facilitate faster inference. 
Step 1 can be decomposed into two steps: computing the saliency map and locating the saliency features from the saliency map. 
The majority of the overhead by \sysname is from computing the saliency map using SmoothGrad. 
The overhead of using SmoothGrad with 15-iteration implementation (i.e., 15 random examples for computing the average gradients) is 340\emph{ms}, and the total overhead by \sysname is 345.7\emph{ms}. 
Nevertheless, the overhead in generating saliency map can be optimized by using more efficient saliency map methods, such as \cite{mundhenk2019efficient}, which can also accurately generate the salinecy maps (hence not affecting \sysname detection efficacy) but with a speedup of 1456\emph{x} over SmoothGrad~\cite{mundhenk2019efficient} (this is because \cite{mundhenk2019efficient} requires only a single forward pass through a few of the layers in a network; while SmoothGrad requires multiple forward and backward pass that are more time-consuming). 
Hence, the overhead by \sysname can be reduced to 5.93\emph{ms} (estimated), while an inference on the undefended model takes 5.56\emph{ms}, which translates to a 6.7\% overhead by \sysname.

\emph{Attack mitigation.} 
\sysname involves using a GAN to recover the uncorrupted examples from adversarial examples and prediction comparison for attack mitigation. 
Note that this process is activated only \emph{after} an attack is detected, hence its overhead is also dependent on the \emph{ratio of adversarial examples in all the test inputs}. For this reason, we plot the overhead under different ratios of adversarial examples, and we show the results in Fig.~\ref{fig:overhead}. 

When the attack ratio is below 1\%, the mitigation overhead by \sysname (including both performing GAN-based recovery and prediction comparison) took less than 7.6\emph{ms} while inference on the undefended model took 5.56\emph{ms}. 
As the attack ratio increases, \sysname incurs higher overhead as it needs to perform more mitigation task on the increasing amount of adversarial examples.

%%%%%%%%%%%%%%%%%%%%%%%%%%%%%%%%%%%%%%%%%%%%%%%%%%%%%%%%%%%%%%%%%%%%%%%%%%%%%%%%
\end{document}